\RequirePackage{pgfplots} 
\documentclass[sn-mathphys,iicol]{sn-jnl}

\pdfoutput=1
\usepackage{graphicx}
\usepackage[lofdepth]{subfig}
\pgfplotsset{compat = newest}
\usepackage{mathtools}
\usepackage[nameinlink]{cleveref}
\usepackage[export]{adjustbox}
\usepackage{siunitx}

\newcommand{\cf}{{cf.~}}
\newcommand{\phys}{{\text{phys}}}
\newcommand{\veps}{\varepsilon}

\newcommand\ie{{i.e.~}}
\newcommand{\bOmp}[2]{b_{\Om_+} \left ({#1},\,{#2}\right )}
\newcommand{\cOmp}[2]{c_{\Om_+} \left ({#1},\,{#2}\right )}

\newcommand{\eg}{e.g.,~}
\newcommand{\wrt}{\text{ with respect to~}}
\newcommand{\tot}{{\text{tot}}}

\newcommand{\nel}{{n_\mathrm{el}}}
\newcommand{\RR}{\mathbb{R}}
\newcommand{\FF}{\mathbb{F}}
\newcommand{\cS}{\mathbb{S}}
\newcommand{\rhobarm}{\bar{\rho}_m}
\newcommand{\Img}{\rm Im}

\newcommand{\Om}{\Omega}
\newcommand{\dlt}{\delta}
\newcommand{\aOm}[2]{a_{\Om} \left ({#1},\,{#2}\right )}
\newcommand{\bOm}[2]{b_{\Om} \left ({#1},\,{#2}\right )}
\newcommand{\cOm}[2]{c_{\Om} \left ({#1},\,{#2}\right )}
\newcommand{\intY}{\sim \kern-1.2em \int}
\newcommand{\intYs}{\smallsim \kern-.75em \int}
\newcommand{\smallsim}{\ensuremath \raisebox{.15em}{{$\scriptstyle\sim$}}}
\newcommand{\aYm}[2]{a_{Y}^m \left ({#1},\,{#2}\right )}
\newcommand{\Dop}{{{\rm I} \kern-0.2em{\rm D}}}
\newcommand{\eeb}[1]{\eb({#1})}
\newcommand{\eeby}[1]{\eb_y({#1})}
\newcommand{\Hdb}{{\bf{H}}^1}
\newcommand{\Hpdb}{{\bf{H}}_\#^1}
\newcommand{\Pibf}{{\mbox{\boldmath$\Pi$\unboldmath}}}
\newcommand{\omegabf}{{{\mbox{\boldmath$\omega$\unboldmath}}}}
\newcommand{\dSy}{\,\mathrm{dS}_{y}}
\newcommand{\mx}{{[m]}}
\newcommand{\psibf}{{\mbox{\boldmath$\psi$\unboldmath}}}
\newcommand{\pd}{\partial}
\newcommand{\vphi}{\varphi}
\newcommand{\Lambdabf}{\mbox{\boldmath$\Lambda$\unboldmath}}
\newcommand{\vthetabf}{\mbox{\boldmath$\vartheta$\unboldmath}}

\newcommand{\ol}[1]{\overline{#1}}
\newcommand{\dvg}{\mbox{\rm div}}

\newcommand{\Aop}{{{\rm A} \kern-0.6em{\rm A}}}
\newcommand{\Hop}{{{\rm I} \kern-0.2em{\rm H}}}

\newcommand{\ab}{{\boldsymbol{a}}}
\newcommand{\eb}{{\boldsymbol{e}}}
\newcommand{\fb}{{\boldsymbol{f}}}
\newcommand{\gb}{{\boldsymbol{g}}}
\newcommand{\nb}{{\boldsymbol{n}}}
\newcommand{\ub}{{\boldsymbol{u}}}

\newcommand{\vb}{{\boldsymbol{v}}}
\newcommand{\wb}{{\boldsymbol{w}}}
\newcommand{\zb}{{\boldsymbol{z}}}
\newcommand{\Bb}{{\boldsymbol{B}}}
\newcommand{\Cb}{{\boldsymbol{C}}}
\newcommand{\Ib}{{\boldsymbol{I}}}
\newcommand{\Kb}{{\boldsymbol{K}}}
\newcommand{\Qb}{{\boldsymbol{Q}}}
\newcommand{\Rb}{{\boldsymbol{R}}}
\newcommand{\Wb}{{\boldsymbol{W}}}
\newcommand{\balpha}{\boldsymbol{\alpha}}

\newcommand{\sigmabf}{{{\mbox{\boldmath$\sigma$\unboldmath}}}}

\newcommand{\Uad}{U_\text{ad}}

\newcommand{\lambrho}{\lambda_\rho}
\newcommand{\LambXi}{\Lambda_\Xi}
\newcommand{\Lambg}{\Lambda_g}
\newcommand{\Jglobtilde}{\Jcaltilde_\mathrm{glob}}
\newcommand{\Jvol}{\Jcal_\mathrm{vol}}
\newcommand{\Jdiff}{\Jcal_\mathrm{diff}}
\newcommand{\Jsep}{\Jcal_\mathrm{sep}}
\newcommand{\Jsepe}{\Jcal_{\mathrm{sep},e}}
\newcommand{\Jvoltilde}{\Jcaltilde_\mathrm{vol}}
\newcommand{\Jreg}{\Jcal_\mathrm{reg}}
\newcommand{\Jregtilde}{\Jcaltilde_{\mathrm{reg},e}}
\newcommand{\Jphys}{\Jcal_\mathrm{phys}}
\newcommand{\Jphystilde}{\Jcaltilde_\mathrm{phys}}
\newcommand{\Rtilde}{\widetilde{R}_e}
\newcommand{\Agrid}{A^\text{grid}}
\newcommand{\ndof}{n_{\text{dof}}}
\newcommand{\Jcaltilde}{\widetilde{\Jcal}}

\newcommand{\Htilde}{\tilde{H}}
\newcommand{\Hcaltilde}{\tilde{\Hcal}}
\newcommand{\Hopd}{{{\bf I} \kern-0.2em{\bf H}}}
\newcommand{\Dopd}{{{\bf I} \kern-0.2em{\bf D}}}

\newcommand{\ubm}{\boldsymbol{\mathrm{u}}}
\newcommand{\pbm}{\boldsymbol{\mathrm{p}}}
\newcommand{\zbm}{\boldsymbol{\mathrm{z}}}

\newcommand{\Ecal}{\mathcal{E}}
\newcommand{\Fcal}{\mathcal{F}}
\newcommand{\Hcal}{\mathcal{H}}
\newcommand{\Jcal}{\mathcal{J}}
\newcommand{\Lcal}{\mathcal{L}}
\newcommand{\Scal}{\mathcal{S}}
\newcommand{\Tcal}{\mathcal{T}}
\newcommand{\Ycal}{\mathcal{Y}}
\newcommand{\Wcal}{\mathcal{W}}
\newcommand{\Zcal}{\mathcal{Z}}

\newtheorem{remark}{\it Remark\/}

\jyear{2023}

\begin{document}

\title[Sequential Global Programming Applied to Fluid-saturated Porous Media]{A Sequential Global Programming Approach for Two-scale Optimization of Homogenized Multiphysics Problems with Application to Biot Porous Media}

\author*[1]{\fnm{Bich Ngoc} \sur{Vu}}\email{bich.ngoc.vu@fau.de}
\author[2]{\fnm{Vladimir} \sur{{Luke\v{s}}}}\email{vlukes@kme.zcu.cz}
\author[1]{\fnm{Michael} \sur{Stingl}}\email{michael.stingl@fau.de}
\author*[2]{\fnm{Eduard} \sur{Rohan}}\email{rohan@kme.zcu.cz}

\affil[1]{\orgdiv{Competence Unit for Scientific Computing}, \orgname{Friedrich-Alexander-Universit\"at Erlangen-N\"urnberg}, \orgaddress{\street{Martenstrasse 5a}, \city{Erlangen}, \postcode{91058}, \country{Germany}}}
\affil[2]{\orgdiv{Department of Mechanics \&  NTIS New Technologies for Information Society}, \orgname{University of West Bohemia in Pilsen}, \orgaddress{\street{Univerzitn\'\i~22}, \city{Plze\v{n}}, \postcode{30614}, \country{Czech Republic}}}

\abstract{
  We present a new approach and an algorithm for optimizing the material configuration and behaviour of a fluid saturated porous medium in a two-scale setting. The state problem is governed by the Biot model describing the fluid-structure interaction in homogenized poroelastic structures. However, the approach is widely applicable to multiphysics problems involving several macroscopic fields where homogenization provides the relationship between the microconfigurations and the macroscopic mathematical model.
  The optimization variables describe the local microstructure design by virtue of the pore shape which determines the effective medium properties -- the material coefficients -- computed by the homogenization method. The main idea of the numerical optimization strategy consists in a) employing a precomputed database of the material coefficients associated to the geometric parameters and b) applying the sequential global programming (SGP) method for solving the problem of macroscopically optimized distribution of material coefficients. Although there are similarities with the free material optimization (FMO) approach, only effective material coefficients are considered admissible, for which a well-defined set of corresponding configurable microstructures exist. Due to the flexibility of the SGP approach, different types of microstructures with fully independent parametrizations can easily be handled. The efficiency of the concept is demonstrated by a series of numerical experiments. We show that the SGP method can handle simultaneously multiple types of microstructures with nontrivial parametrizations using a considerably low and stable number of state problems to be solved.
}

\keywords{
multi-material optimization;  sequential global programming; homogenization; Biot model; poroelasticity; sensitivity analysis
}

\maketitle


\section{Introduction}\label{sec-Introduction}

The design of fluid-saturated poroelastic media (FSPM) present a gradually increasing topic of research interest due to its mathematical complexity and a great application potential. Although the theory of FSPM has been developed in the context of geomechanics and civil engineering, nowadays theses types of materials are abundant in many engineering applications. A convenient  design of microstructures can provide a metamaterial property related to controllable fluid transport, or elasticity. In particular, soft robots can be designed as inflatable porous structures generating a motion and force due to variable fluid content, \eg \cite{Andreasen-Sigmund-2013}. To this aim, the behaviour of the fluid-saturated porous materials is described by the Biot model \cite{Biot1957}, within the small strain theory, which was postulated using a phenomenological approach. The homogenization method enabled the derivation of the quasistatic Biot's equations \cite{Burridge-Keller-1982}. Since then, a number of works extended the results for the dynamic case, which is important for treating wave propagations, see \eg \cite{Rohan-Naili-ZAMP2020}. As an extension beyond the linear theory, a modified Biot model with strain-dependent poroelastic and permeability coefficients was proposed in \cite{Rohan-Lukes-2015}.

Topology optimization of microstructures constituting the FSPM was treated in
\cite{Andreasen_2012} and \cite{Andreasen-Sigmund-2013}. Therein, the fluid-structure interaction problem was handled in the homogenization framework and an approximation towards computational simplification was proposed.

In this paper, we aim at a two-scale approach optimization allowing for a spatial grading of the microstructure design. 
Two-scale optimization problems have been already extensively discussed in literature before. The whole idea started with the seminal paper of Bends{\o}e and Kikuchi \cite{BendsoeKikuchi}, in which the following concept was suggested: for a given parametrization of the unit cell, carry out the homogenization procedure on a fixed parameter grid in a preprocessing step. Then, in every step of the optimization, first retrieve, for each design element, (approximate) effective material coefficients by interpolation. Next, plug these coefficients into the state equation, solve the latter and evaluate the cost. The other way round, sensitivities are computed by the chain rule, i.e. first differentiate the quantity of interest \wrt the material coefficients and then differentiate the material coefficients \wrt to the design parametrization. This procedure opens the way for the application of any suitable gradient based optimization solver, like, \eg OCM \cite{sigmund99}, MMA \cite{Svanberg-MMA-1987} or SnOpt \cite{Gill-Snopt-2002}, to name only those, which are most prominently used in structural topology and material optimization.

While this concept essentially carries over to other classes of problems, as it is done by \cite{das2020,zhou2021,chen2023} for thermomechanical settings, we opted to follow a slightly different avenue in this paper. There are several reasons: First, the concept depends, by its nature, to a large extent on the chosen parametrization. If the parameters enter the homogenized properties in a substantially non-convex way (as it is the case, if, \eg rotations of the base cells are allowed), many local minima might be introduced and additional measures must be taken to avoid getting trapped in one of them. Second, it is not easy to extend the original concept with respect to the use of completely independent types of unit cells, either characterized by different geometries or material configurations. In this case, specifying a smooth parametrization is non-trivial. The typical idea would be to first introduce an independent parametrization for either cell types (for example using sizing variables) and then add on top a smooth interpolation scheme for the effective tensors as used, for instance, in multi-material optimization (see \cite{hvejsel2011}). The problems with that is however, that the second level of interpolation introduces material coefficients, for which typically \emph{no} interpretation in terms of a microstructure exists. Thus, an additional penalization strategy is required, which ensures that those unphysical choices do not remain in the optimal solution. Such an approach was successfully demonstrated in the recent work \cite{YPSILANTIS2022106859}. In another recent article, \cite{LIU2023116485} chose two unit cell types, described via level-set functions, such that the mixture of their geometric parameters can be directly interpreted as a third unit cell type. \cite{PIZZOLATO2019112552} also opted for level-set functions to describe the geometry of the microstructures. But, with respect to the handling of multiple material classes, the authors defined floating patches, where each patch is a subdomain of the design domain and only occupied by one microstructure type. Then, the layout of these patches are optimized on the macroscopic level and their overlaps are combined via a differentiable maximum operator.

In our paper, we describe, how these disadvantages can be circumvented using the SGP concept. The basic idea has been already introduced in \cite{Semmler-SIAM-2018} and is now generalized to a multiphysics, two-scale setting. This involves an extension of an MMA-type block-separable model function (see \cite{stingl-siam-2009}) to the poroelastic setting, a split of the computations into an \emph{offline} and an \emph{online} phase, which is particularly suited for homogenization based problems, and a numerical solution scheme for the nearly global optimization of block-separable subproblems. We would like to note here that the term \emph{block-separable} implies that the minimization can be carried out separately for each design element, however a design element itself can be described by multiple design degrees of freedom. For a further motivation of the SGP method, we refer to the first paragraph in \cref{sec:sgp}.
Here, we just like to add that, in the whole optimization process, two different types of sensitivities are relevant. First, there are the sensitivities of constraint or cost functions \wrt the effective material coefficients. These constitute a substantial ingredient of the block-separable model used in the heart of the SGP method. Second, there are the sensitivities of the material coefficients \wrt the chosen parametrization. In the context of the suggested two-scale SGP framework, the latter ones are not strictly required, but can help to come up with an improved interpolation model used in the offline phase. In any case, the derivation of sensitivities presented in this paper, for the particular context of fluid saturated porous media, relies on derivations in \cite{Huebner-Solid-2019}, where also the sensitivity of the homogenized coefficients were reported, see also \cite{Rohan-Lukes-2015}.

Finally, we would like to comment on the generality of the presented approach.
Although the SGP concept outlined in our paper can be applied to a large range of multiphysics two-scale material optimization problems, the Biot model of fluid saturated porous media
provides an ideal test bed for the method. This is for several reasons: first, the physical coupling is non-trivial. Second, it is very natural to set up competing objective functions, such as the structural compliance on the one hand and the enhanced fluid flow through an outflow boundary, on the other hand. And third, configurable types of microstructures supporting either the first or the second goal can be deduced in a straightforward manner.

The structure of the remainder of this paper is as follows: In \cref{sec-twoscale} all ingredients of the two-scale problem are described. To these belong a brief repetition of the constitutive laws for the Biot model (\cref{sec-biot}), the poroelastic state problem in variational form (\cref{sec:state-problem}), a generic sketch of the two-scale problem constrained by the poroelasticity equations (\cref{sec:two-scale-opt-problem}) and an adjoint analysis providing sensitivities with respect to effective material coefficients, as used later by the SGP method (\cref{sec-sensitivity}). Finally, two types microstructures are suggested in form of configurable unit cells (\cref{sec:design_params}). In \cref{sec:sgp} the SGP concept for the solution of two-scale optimization problems is introduced in greater detail. For this, the two-scale problem is discretized and extended for the use of multiple types of unit cells (\cref{sec-twoscale-discr}). Then, a separable sequential approximation concept is suggested (\cref{sec:subproblems}) and last the SGP method is presented in an algorithmic form (\cref{sec:SGPalg}). In \cref{sec:numerical-results}, the advantages of the SGP algorithm will be discussed using various types of two-scale problems. 

\section{Formulation of the two-scale optimization problem} \label{sec-twoscale}
In this section, we explain our optimization strategy. Although it can be applied to similar problems involving several physical fields or multiphysics problems, in this paper, we consider the fluid saturated porous media represented by the Biot model which can be derived using the homogenization of the fluid-structure interaction problem restricted to small deformation kinematics, see \eg  \cite{Burridge-Keller-1982,Brown2011,Rohan-Naili-Lemaire-CMAT2015}. In the next section we report the homogenization result presented 

\paragraph{Notation}
We employ the following notation. 
Since we deal with a two-scale problem, we distinguish the ``macroscopic'' and ``microscopic'' coordinates, $x$ and $y$, respectively.
We use ${\nabla_x = (\pd_i^x)}$ and ${\nabla_y = (\pd_i^y)}$ when differentiation \wrt coordinate $x$ and $y$ is used, respectively, whereby $\nabla \equiv \nabla_x$. By $\eeb{\ub} = 1/2[(\nabla\ub)^T + \nabla\ub]$, we denote the strain of a vectorial function $\ub$, where the transpose operator is indicated by the superscript ${}^T$.
The Lebesgue spaces of 2nd-power integrable functions on
an open bounded domain $D\subset \RR^3$ is denoted by $L^2(D)$, the Sobolev space $\Wb^{1,2}(D)$
of the square integrable vector-valued functions on $D$ including the first order
generalized derivative, is abbreviated by $\Hdb(D)$.  Further,
$\Hpdb(Y_m)$ is the Sobolev space of vector-valued Y-periodic
functions (the subscript $\#$).

\subsection{The homogenized Biot -- Darcy model}\label{sec-biot}
We report the homogenization result presented \eg in \cite{Rohan-Lukes-2015}, \cf \cite{Huebner-Solid-2019}, where the problem of locally optimized microstructures has been described.
The homogenized model of the porous elastic medium incorporates local problems for characteristic responses which are employed to compute the effective material coefficients of the Biot model. 

The local problems specified below, related to the homogenized model,
are defined at the microscopic representative unit cell ${Y =
\Pi_{i=1}^3]0,\ell_i[ \subset \RR^3}$. which splits into the solid part
occupying domain $Y_m$ and the complementary channel part $Y_c$. Thus,
\begin{align}\label{eq-6}
Y   &= Y_m  \cup Y_c  \cup \Gamma_Y \;,\quad \nonumber\\
Y_c   &= Y \setminus Y_m  \;,\quad \nonumber\\
\Gamma_Y   &= \ol{Y_m } \cap \ol{Y_c }\;,
\end{align}
where by $\ol{Y_d }$ for $d=m,c$, we denote the closure of the open bounded domain $Y_d$. By $\intYs_{Y_d} = \vert Y \vert^{-1}\int_{Y_d}$, with
$Y_d\subset \ol{Y}$ for $d=m,c$, we denote the local average ($\vert Y\vert$ is the volume of domain  $Y$). Obviously, the unit volume $\vert Y\vert=1$ can always be chosen. We employ the usual elasticity bilinear form, involving two vector fields $\wb$ and $\vb$, that reads
\begin{equation}
\aYm{\wb}{\vb} = \intY_{Y_m} (\Dop \eeby{\wb}) : \eeby{\vb}\;,
\end{equation}
where $\Dop = (D_{ijkl})$ is the elasticity tensor satisfying the usual symmetries, $D_{ijkl} = D_{klij} = D_{jikl}$, and $\eeby{\vb} = \frac{1}{2}(\nabla_y \vb + (\nabla_y \vb)^T)$ is the linear strain tensor associated with the displacement field $\vb$.

In what follows, by the microstructure $\Ycal(x)$, we mean the decomposition \cref{eq-6} of the representative cell $Y$ and the material properties, as represented by the elasticity $\Dop$ only in our case.
If the structure is perfectly periodic, 
microstructures $\Ycal \equiv \Ycal(x)$ are independent of
the macroscopic position $x \in \Om$. Otherwise, the local problems
must be considered at any macroscopic position, i.e. for almost any
$x \in \Om$, see \eg \cite{Brown2011} in the context of slowly varying ``quasi-periodic'' microstructures.
It should be pointed out, that this issue is of a special importance when dealing with homogenization-based material design optimization; as will be explained below, a regularization is required to control the design variation within $\Om$.

The local microstructural response is obtained by solving the following decoupled problems:
\begin{itemize}
    \item Find
    $\omegabf^{ij}\in \Hpdb(Y_m)$ for any $i,j = 1,2,3$
    satisfying
    \begin{equation}\label{eq-h5a}
    \begin{split}
    \aYm{\omegabf^{ij} + \Pibf^{ij}}{\vb} & = 0\;,\; \forall \vb \in  \Hpdb(Y_m)\;,
    \end{split}
    \end{equation}
   where $\Pibf^{ij} = (\Pi_k^{ij})$, $i,j,k   = 1,2,3$ with components $\Pi_k^{ij} = y_j\delta_{ik}$.
    \item Find
    $\omegabf^P \in \Hpdb(Y_m)$
    satisfying
    \begin{equation}\label{eq-h5b}
    \begin{split}
    \aYm{\omegabf^P}{\vb} & = \intY_{\Gamma_Y} \vb\cdot \nb^\mx \dSy, \;
    \forall \vb \in  \Hpdb(Y_m) \;.
    \end{split}
    \end{equation}
    \item Find $(\psibf^i,\pi^i) \in \Hpdb(Y_c) \times L^2(Y_c)$ for $i = 1,2,3$ such that
    \begin{equation}\label{eq-S3}
    \begin{split}
    \int_{Y_c} \nabla_y \psibf^k: \nabla_y \vb -
    \int_{Y_c} \pi^k \nabla\cdot \vb  &= \int_{Y_c} v_k\;,\\ 
    \int_{Y_c} q \nabla_y \cdot \psibf^k & = 0\;, \\ 
    \end{split}
    \end{equation}
\end{itemize}
$\forall \vb \in \Hpdb(Y_c)$ and $\forall q \in L^2(Y_c)$.

Effective material properties of the homogenized deformable fluid-saturated porous medium are described in terms of  homogenized poroelastic coefficients: the drained elasticity $\Aop$, the stress coupling $\Cb$ and the compressibility $N$, all being related to the solid skeleton. All these coefficients including the intrinsinc hydraulic permeability $\Kb$ are computed using the characteristic microscopic responses \cref{eq-h5a,eq-h5b,eq-S3} substituted in following expressions:
\begin{equation}\label{eq-h8}
\begin{split}
A_{ijkl} = \aYm{\omegabf^{ij} + \Pibf^{ij}}{\omegabf^{kl} + \Pibf^{kl}}\;,\quad \\
C_{ij} =  -\intY_{Y_m}\dvg_y \omegabf^{ij} = \aYm{\omegabf^P}{\Pibf^{ij}}\;,\\
N  =  \aYm{\omegabf^P}{\omegabf^P} = \intY_{\Gamma_Y} \omegabf^P\cdot \nb \dSy\;,\\
K_{ij}  = \intY_{Y_c} \psi_i^j =  \intY_{Y_c} \nabla_y \psibf^i : \nabla_y \psibf^i \;.
\end{split}
\end{equation}
Obviously, the tensors $\Aop = (A_{ijkl} )$, $\Cb = (C_{ij} )$ and
$\Kb = (K_{ij} )$ are symmetric, $\Aop$ adheres all the symmetries of
$\Dop$; moreover $\Aop$ is positive definite and $N > 0$. The
hydraulic permeability $\Kb$ is, in general, positive semi-definite.
It is positive definite whenever the channels constitute a simply connected domain generated as the periodic lattice by $Y_c$; 
for this, denoting by $\Gamma_Y^k\subset \pd Y$, $k=1,\dots,6$ the faces of $Y$, it must hold that $\Gamma_Y^k\cap \pd Y_c \not = \emptyset$ for all $k=1,\dots,6$.

\subsection*{Coupled flow deformation problem}
The Biot--Darcy model of poroelastic media for quasi-static, evolutionary problems
imposed in $\Om$ is constituted by the following equations involving
stress $\sigmabf$, displacement $\ub$,  strain $\eb(\ub)$,  
fluid pressure $p$ and the seepage velocity $\wb$:
\begin{equation}\label{eq-B1}
\begin{split}
-\nabla \cdot \sigmabf  & = \fb^s,\; \quad
\sigmabf  = \Aop \eeb{\ub} - \Bb p,\\
 -\nabla\cdot \wb & = \Bb:\eeb{\dot\ub} +  M \dot p, \\
\wb &= -\frac{\Kb}{\bar\eta} \left (\nabla p - \fb^f \right),
\end{split}
\end{equation}
where the homogenized coefficients are given by \cref{eq-h8} and
\begin{equation}\label{eq-B1a}
\begin{split}
\Bb & := \Cb + \phi \Ib\;,\\
M & := N + \phi \gamma\;.
\end{split}
\end{equation}
Above, $\bar\eta$ is the relative fluid viscosity, $\gamma$ is the fluid compressibility and  $\phi = \vert Y_c\vert / \vert Y \vert$ is the porosity (volume fraction of the fluid-filled channels). The effective volume forces in \cref{eq-B1}, acting in the solid and fluid phases, are denoted by $\fb^s$ and $\fb^f$, respectively.
It is important to note that $\bar\eta = \eta^\phys/\veps_0^2$ is
defined for a given fluid ($\eta^\phys$) and microstructures scale: $\veps_0 = \ell_0/L$ where $L$ is a characteristic macroscopic length, and $\ell_0$ is the characteristic microstructure size, typically given by the ``pore diameter''. Thus, for a given fluid, the effective permeability $\Kb/\bar\eta$ is proportional to $\veps_0^2$, \ie reflecting the microstructure size. In contrast, all other coefficients are scale-independent (when the scale separation holds, \ie $\veps_0$ being  small enough).

\begin{remark}\label{rem1}
 In this paper, we only consider steady state problems for the Biot medium, such that all time derivatives in \cref{eq-B1} vanish. Consequently, the Biot compressibility $M$ is not involved, as far as the porous phase, generated as a periodic lattice by channels $Y_c$, is connected. For any microstructure with disconnected pores, such that $\ol{Y_c} \subset Y$, thus, $Y_c$ constitute one, or more inclusions with one cell $Y$, see \cite{Rohan-Naili-Lemaire-CMAT2015}, the permeability vanishes. Then, the time integration in \cref{eq-B1} leads to the mass conservation equation in the form $\Bb:\eeb{\ub} +  M p = 0$, assuming an undeformed initial configuration with the zero pressure in the inclusions.
In the optimization problem, besides microstructures with nondegenerate permeabilities,  we shall consider  also microstructures with spherical, thus, disconnected pores, constituting impermeable material. For this case, one can choose either fluid filled pores, or empty pores; the only difference is the use of the so-called undrained material elasticity, $\Aop_U = \Aop + M^{-1}\Bb\otimes\Bb$, or the elasticity $\Aop$ describing effective elasticity of the ``drained'' skeleton, with empty pores. 
\end{remark}

\subsection{State problem formulation}\label{sec:state-problem}
Let $\Om \subset \RR^3$ be an open bounded domain. Its boundary $\pd\Om$ splits, as follows:
$\pd\Om = \Gamma_D \cup \Gamma_N$ and also $\pd\Om = \Gamma_p \cup \Gamma_w$, where $\Gamma_D \cap \Gamma_N = \emptyset$ and $\Gamma_p \cap \Gamma_w = \emptyset$. Assume $\Gamma_p$ consists of two disconnected, non-overlapping  parts $\Gamma_p^k$, $k = 1,2$,
$\Gamma_p = \Gamma_p^1\cup \Gamma_p^2$, and $\Gamma_p^1\cap \Gamma_p^2 = \emptyset$.

We consider the steady state problems for the linear Biot continuum occupying domain $\Om$.
The poroelastic material parameters and the hydraulic permeability referred to as the homogenized coefficients, in general, are given by the locally defined  microstructures $\Ycal(x)$ which can vary with ${x \in \Om}$.
The two-scale optimization approach proposed in this paper enables to combine microstructures characterized by connected and disconnected pores, the latter characterized by a vanishing permeability. To this aim, the domain $\Om = \Om_0 \cup \Om_+$ is decomposed into in two parts: the permeable $\Om_+$  and the impermeable $\Om_0$, which may not constitute connected domains, being split into more disconnected subparts. Consequently, the interface $\Gamma_+ = \pd\Om_+ \cap \pd\Om_0$ is impermeable.
Regarding the boundary decomposition, we assume that $\Gamma_{p+}^k := \Gamma_p^k\cap\pd\Om_+ \not = \emptyset$, for $k=1,2$, so that the porous structure permits the fluid transport through domain $\Om_+$, if this one connects $\Gamma_{p+}^1$ and $\Gamma_{p+}^2$.

We consider the following macroscopic problem: Given the traction surface forces $\gb$, and pressures $\bar p^k$ on boundaries $\Gamma_p^k$, 
find displacements $\ub$ and the hydraulic pressure $P$ which satisfy
\begin{equation}\label{eq-opg1}
\begin{split}
-\nabla\cdot\left(\Aop \eeb{\ub} - P \Bb\right) & = 0\quad \mbox{ in } \Om\;,\\
\ub & = 0 \quad \mbox{ in }\Gamma_D\;,\\
\left(\Aop \eeb{\ub} - P \Bb\right) \cdot \nb & = \gb \quad \mbox{ in }\Gamma_N\;,
\end{split}
\end{equation}
where $P = 0$ in $\Om_0$. Whereas, in $\Om_+$, $P$ satisfies
\begin{equation}\label{eq-opg2}
\begin{split}
-\nabla\cdot\Kb\nabla P & = 0\quad \mbox{ in } \Om_+\;,\\
P & = \bar p^k \quad \mbox{ on } \Gamma_{p+}^k\;,\quad k = 1,2\;,\\
\nb\cdot\Kb\nabla P & = 0 \quad \mbox{ on } \Gamma_w \cup \Gamma_+\;.
\end{split}
\end{equation}

For the steady state problem the set of equations \cref{eq-B1} yields the two problems \cref{eq-opg1} and \cref{eq-opg2} as a decoupled system: first, \cref{eq-opg2} can be solved for $P$, then \cref{eq-opg1} is solved for $\ub$. Moreover, for the considered type of the boundary conditions and since volume forces are not involved, the solutions are independent of the viscosity $\bar\eta$, see \cref{eq-B1}.

Further, we consider an extension of $\bar p^k$ from boundary $\Gamma_p^k$ to the whole domain $\Om$, such that
$\bar p^k = 0$ on $\Gamma_p^l$ (in the sense of traces) for $l\not = k$.
Then $P = p + \sum_k\bar p^k$ in $\Om_+$, such that $p = 0$ on $\Gamma_{p+}$.
Note that $p$ can be simply extended by 0 in $\Om_0$.
For the sake of notational simplicity, we introduce $\bar p=\sum_k\bar p^k$.
By virtue of the Dirichlet boundary conditions for $\ub$ and $p$, we introduce the following spaces:

\begin{equation}\label{eq-opg3}
\begin{split}
V_0 & = \{\vb \in \Hdb(\Om)\, \vert \; \vb = 0 \mbox{ on } \Gamma_D\}\;,\\
Q_0 & = \{q \in L^2(\Om)\cap H^1(\Om_+)\, \vert \; q = 0 \mbox{ on } \Gamma_{p+}\}\;.
\end{split}
\end{equation}
We employ the bilinear forms and the linear functional $g$,
\begin{equation}\label{eq-opg4}
\begin{split}
\aOm{\ub}{\vb} & = \int_\Om (\Aop\eeb{\ub}):\eeb{\vb}\;,\\
\bOmp{p}{\vb} & = \int_{\Om_+} p \Bb:\eeb{\vb}\;,\\
\cOmp{p}{q} & = \int_{\Om_+} \nabla q\cdot \Kb\nabla p\;,\\
g(\vb) & = \int_{\Gamma_N} \gb \cdot \vb\;.
\end{split}
\end{equation}

In order to define the state problem in the context of two-scale optimization, we employ the weak formulation which reads, as follows: Find $\ub\in V_0$ and $p \in Q_0$, such that, for all $\vb \in V_0$ and $q \in Q_0$,
\begin{equation}\label{eq-opg5}
\begin{split}
    \aOm{\ub}{\vb} - \bOmp{p}{\vb} & =  g(\vb) + \bOmp{\bar p}{\vb}, \\ 
    \cOmp{p}{q} & =  -\cOmp{\bar p}{q}. 
\end{split}
\end{equation}
To define $p$ uniquely in $\Om$, $p\equiv 0$ in $\Om_0 = \Om\setminus \Om_+$.
Since the two fields are decoupled, first $p$ is solved from \cref{eq-opg5}$_2$, then
$\ub$ is solved from \cref{eq-opg5}$_1$, where $p$ is already known.
\begin{remark}\label{rem2}
  In the context of the undrained porosity defined by fluid-filled closed pores $Y_c \subset Y$, see Remark~\ref{rem1},
  formulation \cref{eq-opg5} is consistent also with this microstructure class type $\Ycal_0^\square$ with $\Aop_U$ replacing $\Aop$ in the elasticity bilinear form \cref{eq-opg4}$_1$. Pressure is then defined pointwise in $\Om_0$ by $P:= - \Bb:\eeb{\ub}/M$.
\end{remark}

 By $\alpha(x)$ we denote an abstract optimization variable which determines the homogenized coefficients for any position $x\in\Om$. Below we consider $\alpha$ representing several geometrical parameters characterizing microstructures $\Ycal(x)$ of a given type.
Although, in this section, we disregard some particular details related to the treatment of multiple types of $\Ycal$, we bear in mind the existence of two microstructure classes, $\Ycal_+^\square$ and $\Ycal_0^\square$, associated with the pore connectivity type, as discussed above. The ``permeable'' domain $\Om_+$ is occupied by the material given pointwise by $\Ycal(x) \in \Ycal_+^\square$ for all $x\in \Om_+$. Hence, both the subdomains of $\Om$ are defined implicitly by the microstructure type:  $\Om_i$ is the set of $x\in \Om$, such that $\Ycal(x) \in \Ycal_i^\square$, where $i = +,0$.  

In the next section, we shall consider a two-scale optimization problem which is characterized by the following features:
\begin{itemize}
\item Geometrical restrictions are stated in respective definitions of the admissibility designs sets for a chosen type of microstructure. For the sake of brevity, let $A$ be the set of admissible designs, further we consider $\alpha(x) \in A$  for any $x \in \Om$. 
\item We consider multiple optimization criteria which perform as the objective functions, or equality constraints. Without loss of generality, we confine ourselves to the two criteria $\Phi_\alpha(\ub)$ and $\Psi_\alpha(p)$ that are defined, as follows:
\begin{equation}\label{eq-opg6}
\begin{split}
\Phi_\alpha(\ub) & = g(\ub)\;,\\
\Psi_\alpha(p) & = - \int_{\Gamma_p^2}\Kb\nabla (p+ \bar p) \cdot \nb\;.
\end{split}
\end{equation}
While $\Phi_\alpha(\ub)$ expresses the structural compliance, 
criterion function $\Psi_\alpha(p)$ expresses the amount of the fluid flow  through surface $\Gamma_p^2$ due to the pressure difference $\bar p^1 - \bar p^2$, see
the boundary condition  \cref{eq-opg2}$_2$. These two criteria are antagonist: the pore volume reduction leads naturally to stiffening the structure, but reduces the permeability. Hence, for the objective function $\Phi_\alpha$, function $\Psi_\alpha$ serves as a constraint and vice versa.
\end{itemize}

\subsection{Two-scale optimization problem} \label{sec:two-scale-opt-problem}
Here, for the ease of notation, we restrict to one microstructure type only, namely $\Ycal(x) \in \Ycal_+^\square$, so that we may consider $\Om\equiv\Om_+$. Hence, all the bilinear forms in \cref{eq-opg4} are defined by integration in $\Om$.
Later, in \cref{sec:sgp}, we will consider microstructures characterized by different unit cell types of classes $\Ycal_+^\square$ and $\Ycal_0^\square$, however, the formulations introduced below can be adapted easily.

We first define the direct optimization problem to find design $\alpha(\Om)$ that minimizes a cost functional based on the criteria defined in \cref{eq-opg6}. Further, we introduce the set
$
\Tcal = \cS^6 \times \cS^3 \times \cS^3 \times \RR \times \RR 
$ 
and denote by $\Hop = (\Aop,\Bb,\Kb,\rho_m,R) \in \Tcal$ the (local) material parameters involing  the effective (homogenized) material coefficients, the solid part volume $\rho_m = 1-\phi = \vert Y_m \vert/ \vert Y\vert$, and a regularization parameter $R$, which typically depends only on the design.
We note that the dimension of the regularization label $R$ is, for ease of notation, chosen as 1 for now, although later in \cref{sec:two-cells-reg} more general regularization labels are used.
Obviously, $\Hop$ is given uniquely by the local admissible design $\alpha(x)  \in A$, $x \in \Om$, 
whereby for a suitably chosen parametrization,
the admissibility set is given simply by
\begin{equation*}
A = [\underline{\ab},\overline{\ab}] \subset \RR^n.
\end{equation*}
Examples for such parametrizations along with a description of the lower and upper bounds $\underline{\ab},\overline{\ab} \in \RR^n$ are presented in \cref{sec:design_params}.

For a given admissible design $\alpha(\Om)$, the state $\zb = (\ub,p)$ is the solution of \cref{eq-opg5}, where the homogenized coefficients $\Hop(\alpha)$ are given in \cref{eq-h8} using the characteristic responses $\Wb(\alpha):=(\omegabf^{ij},\omegabf^P,\psibf^k,\pi^k)$. $\Wb(\alpha)$ are the solutions of \cref{eq-h5a,eq-h5b,eq-S3}, which depend on $\alpha(x)$ in terms of the microconfigurations $\Ycal(x)$. In this way, mapping $\Scal:\alpha(\Om) \mapsto \zb(\Om)$ introduces the admissible state.

It can be defined by a composition map, $\Scal = \Zcal\circ\Ecal\circ\Wcal$, where  $\Wcal$ represents the resolvents of the characteristic problems imposed on the local microconfigurations, $\Ecal$ provides the  homogenized material, and $\Zcal$ is the resolvent of the macroscopic state problem, so that
\begin{equation}\label{eq-maps}
\begin{split}
  \Wcal:\alpha & \mapsto W\;,\\
  \Ecal:(\alpha, W) & \mapsto \Hop\;,\\
  \Zcal:\Hop(\Om)& \mapsto\zb(\Om)\;.
\end{split}
\end{equation}

Further, we employ the mapping $$\Hcal:\alpha \mapsto \Hop ,$$  such that
$ \Hcal = \Ecal\circ\Wcal $ is the composition map defined for any admissible design $\alpha(x) \in A$, for a.a. $x \in \Om$.

The macroscopic state problem is the implicit form of the mapping $\Zcal: \Hop \mapsto \zb$, such that ${\zb \in S_0 = V_0\times Q_0}$ satisfies
\begin{equation}\label{eq-SP0}
\begin{split}
\vphi_{\Hop}(\zb,\vb) = 0\quad \forall \vb \in S_0\;,
\end{split}
\end{equation}
where $S_0$ is the space of admissible state problem solutions. For the Biot medium problem, \cref{eq-SP0} is identified with \cref{eq-opg5}.

\subsubsection{Direct two-scale optimization problem}
For the given two functions of interest $\Phi$ and $\Psi$, both depending on the material distribution $\Hop(x)$ and the state $\zb(x)$,  the two-scale abstract optimization problem reads:
\begin{equation}\label{eq-opA1}
\begin{split}
  \min_{\alpha \in A}\ & \Phi(\Hop,\zb) + \LambXi \Xi(\Hop) \\
\mbox{ s.t. } & \Psi(\Hop,\zb) = \Psi_0\;,\\
& \zb = \Scal(\alpha),\\
& \Hop = \Hcal(\alpha),\\
& \int_\Om \rho_m \leq \bar \rho_m \vert\Om\vert\;,
\end{split}
\end{equation}
where the term $\Xi(\Hop)$ in the objective is related to the design regularization, namely to parameter $R$, and $\LambXi \in \RR^+$ is a penalty parameter.
Recall the chain mapping ${\Hcal:\alpha(x) \mapsto \Hop(x)}$ for any $x\in \Om$, then $\zb = \Zcal(\Om)$. 
Below, we abbreviate ${\Phi_\alpha(\zb) =: \Phi(\Hcal(\alpha),\zb)}$ and also
${\Psi_\alpha(\zb) =: \Psi(\Hcal(\alpha),\zb)}$. In \cref{eq-opg6}, specific examples relevant for the Biot medium optimization were given. 

Optimization problem \cref{eq-opA1} is associated with the following inf-sup problem, 
\begin{equation}\label{eq-opA2}
\begin{split}
\min_{\alpha \in A} \inf_{\zb \in S_0} \sup_{ \Lambdabf \in \RR^2, \tilde\zb\in S_0}& \Lcal(\alpha, \zb,\Lambdabf, \tilde\zb)\;,
 \end{split}
\end{equation}
with the Lagrangian function,
\begin{equation}\label{eq-opA3}
\begin{split}
\Lcal(\alpha, \zb,\Lambdabf, \tilde\zb) & =
\Lambda_\Phi\Phi_\alpha(\zb) \\
& + \LambXi \Xi(\Hcal(\alpha)) \\
& + \Lambda_\Psi (\Psi_\alpha(\zb)-\ol{\Psi_0}) \\   & +\vphi_{\Hop(\alpha)}(\zb,\tilde\zb)\;,
\end{split}
\end{equation}
where  
$\Lambdabf = ( \Lambda_\Phi,\Lambda_\Psi) \in \RR^2$ are the Lagrange multipliers associated with the objective and constraint functionals $\Phi$ and $\Psi$, and $\tilde\zb\in S_0$ are Lagrange multipliers -- the adjoint variables --- associated with the constraints of the problem \cref{eq-opA1}.

For a while, we may consider material coefficients
$\Hop$ as the optimization variables (although they are parameterized by $\alpha \in A$). Further, let us  assume
a given value $\Lambdabf \in \RR^2$; note that the entries of $\Lambdabf$ can be positive or negative depending on the desired flow augmentation, or reduction. In the numerical examples, we chose $\Lambda_\Phi > 0$, whereas $\Lambda_\Psi < 0$ indicates the constraint effect of $ \Psi$ relative to $\Phi$.  Upon
denoting by $\Img(\Hcal) =\Hcal(A)$, the image space of all admissible designs, and defining
\begin{align*} \Uad = & \left\{\Hop \in L^\infty(\Omega;\Tcal) \,\vert\, \Hop(x) \in \Img(\Hcal) \, \right.\\
& \left. \text{ for a.a. } x\in \Omega \right\},
\end{align*}
the optimization
problem \cref{eq-opA1} can be rephrased as the two-criteria minimization problem,
\begin{equation}\label{eq-opH1}
\begin{split}
 \min_{
\begin{array}{c}
\Hop \in \Uad
\end{array}
} \ &  \Fcal(\Hop,\zb)\;, \\
\mbox{ s.t. } & \zb = \Zcal(\Hop)\;\\
& \int_\Om \rho_m \leq \bar \rho_m \vert\Om\vert\;,
\end{split}
\end{equation}
where
\begin{align*}
\Fcal(\Hop,\zb) &= \Lambda_{\Phi} \Phi(\Hop,\zb) + \Lambda_{\Psi}\Psi(\Hop,\zb) + \LambXi \Xi(\Hop) \;.
\end{align*}

For the Biot medium optimization, where the two criterion functions $\Phi_\alpha$ and $\Psi_\alpha$ are given in \cref{eq-opg6}, the Lagrangian function attains the form
\begin{equation}\label{eq-opg8}
\begin{split}
&\Lcal(\alpha, (\ub,p),\Lambdabf, (\tilde\vb,\tilde q)) \\
  & = \Lambda_\Phi\Phi_\alpha(\ub) + \Lambda_\Psi (\Psi_\alpha(p)-\ol{\Psi_0}) + \Lambda_\Xi \Xi_\alpha(\Hop)\\
  & \quad + \aOm{\ub}{\tilde\vb} - \bOm{p+\bar p}{\tilde\vb} \\
& \quad - g(\tilde\vb) + \cOm{p+\bar p}{\tilde q}\;.
\end{split}
\end{equation}

\subsection{Adjoint responses and the sensitivity analysis} \label{sec-sensitivity}
In this section, we provide details concerning the sensitivity analysis employed in the preceding section.
We consider $\alpha$ to represent a general optimization variable which is related to the effective medium parameters $\Hop$.
It is worth to note that one may also consider $\alpha \equiv \Hop$ in the context of the free material optimization (FMO).

To obtain the adjoint equation,  we consider  the optimality condition for $(\ub,p)$. Thus, from \cref{eq-opg8} it follows that
\begin{equation}\label{eq-opg10}
\begin{split}
&\dlt_{(\ub,p)}\Lcal(\alpha, (\ub,p),\Lambdabf, (\tilde\vb,\tilde q))\circ (\vb,q) \\
& =
\Lambda_\Phi\dlt_\ub \Phi_\alpha(\ub;\vb) + \Lambda_\Psi \dlt_p\Psi_\alpha(p; q)  \\
& \quad + \aOm{\vb}{\tilde\vb} - \bOm{q}{\tilde\vb} + \cOm{q}{\tilde q} \;,
\end{split}
\end{equation}
where
\begin{equation}\label{eq-opg11a}
\begin{split}
\dlt_\ub\Phi_\alpha(\ub;\vb) &= g(\vb), \\
\dlt_p\Psi_\alpha(p; q) &= -\int_{\Gamma_p^2} \Kb\nabla q \cdot\nb.
\end{split}
\end{equation}
To avoid computation of the gradient $\nabla q$ on ${\Gamma_p^2\subset \pd \Om}$, we consider $\tilde p \in H^1(\Om)$ such that
$\tilde p = 0$ on $\Gamma\setminus\Gamma_p^2$, while $\tilde p = 1$ on $\Gamma_p^2$, then it is easy to see that
\begin{equation}\label{eq-opg12}
\begin{split}
-\Psi_\alpha(p) & = r(p) := \cOm{p+\bar p}{\tilde p}\;,\\
-\dlt_p\Psi_\alpha(p; q) & = \dlt_p r(p;q) = \cOm{q}{\tilde p}\;.
\end{split}
\end{equation}
The optimality conditions \cref{eq-opg10}, related to the state admissibility, yield
the adjoint state ${(\tilde\vb,\tilde q) \in V_0 \times Q_0}$ which satisfies the following identities:
\begin{align}\label{eq-opg11b}
\forall \vb \in V_0&:& \aOm{\vb}{\tilde\vb} & = - \Lambda_\Phi\dlt_\ub\Phi_\alpha(\ub;\vb)\;, \nonumber\\
\forall q \in Q_0&:& \cOm{q}{\tilde q} & = \bOm{q}{\tilde\vb} - \Lambda_\Psi\dlt_p\Psi_\alpha(p; q). 
\end{align}
These equations can be rewritten using \cref{eq-opg11a} and \cref{eq-opg12}, as follows for all $(\tilde\vb,\tilde q) \in V_0 \times Q_0$:
\begin{align}\label{eq-opg11}
\forall \vb \in V_0&:& \aOm{\vb}{\tilde\vb} & = -  \Lambda_\Phi g(\vb)\quad \;, \nonumber\\
\forall q \in Q_0&:& \cOm{q}{\tilde q} & = \bOm{q}{\tilde\vb} + \Lambda_\Psi \cOm{q}{\tilde p}. 
\end{align}
To allow for the independence of the state adjoint on $\Lambdabf$, we define the split
\begin{equation}\label{eq-opg11e}
\begin{split}
  \tilde\vb & = \Lambda_\Phi\tilde\vthetabf\;,\\
  \tilde q & = \Lambda_\Phi\tilde q_1 + \Lambda_\Psi \tilde q_2\;,
\end{split}
\end{equation}
where $\tilde\vthetabf$ and $\tilde q_k$, $k = 1,2$ satisfy for all ${(\tilde\vb,\tilde q,\tilde q) \in V_0 \times Q_0}$ 
\begin{equation}\label{eq-opg11f}
\begin{split}
\forall \vb \in V_0:\quad \aOm{\vb}{\tilde\vthetabf} & = -  g(\vb),\\
\forall q \in Q_0:\quad \cOm{q}{\tilde q_1} & = \bOm{q}{\tilde\vthetabf},\\
\forall q \in Q_0:\quad \cOm{q}{\tilde q_2} & =  \cOm{q}{\tilde p}.
\end{split}
\end{equation}
We can compute the total variation of the Lagrangian with
\begin{equation}\label{eq-opg13}
\begin{split}
\dlt_\alpha^\tot \Lcal & = \Lambda_\Phi\dlt_\ub g(\ub;\dlt_\alpha\ub) - \Lambda_\Psi \dlt_p r(p;\dlt_\alpha p) \\
& \quad + \Lambda_\Phi\dlt_\alpha g(\ub) - \Lambda_\Psi  \dlt_\alpha r(p)  + \Lambda_\Xi \dlt_\alpha\Xi_\alpha(\Hop) \\
& \quad + \aOm{\dlt_\alpha\ub}{\tilde\vb} - \bOm{\dlt_\alpha p}{\tilde\vb}
+\cOm{\dlt_\alpha p}{\tilde q}\\
& \quad + \dlt_\alpha\aOm{\ub}{\tilde\vb}-\dlt_\alpha\bOm{p+\bar p}{\tilde\vb} \\
& \quad +\dlt_\alpha\cOm{p+\bar p}{\tilde q}\;.
\end{split}
\end{equation}
If the pair $(\ub,p)$  solves the state problem and $(\tilde\vb,\tilde q)$ is its adjoint state,
\cref{eq-opg13} is equivalent to the following expression:
\begin{align}\label{eq-opg13a}
\dlt_\alpha^\tot \Lcal & = \Lambda_\Phi\dlt_\alpha g(\ub) - \Lambda_\Psi  \dlt_\alpha r(p) + \Lambda_\Xi \dlt_\alpha\Xi_\alpha(\Hop) \nonumber\\
& \quad \quad + \dlt_\alpha\aOm{\ub}{\tilde\vb}-
\dlt_\alpha\bOm{p+\bar p}{\tilde\vb} \nonumber\\
&\quad \quad+\dlt_\alpha\cOm{p+\bar p}{\tilde q}\;.
\end{align}
Above, the shape derivatives $\dlt_\alpha$ of the bilinear forms can be rewritten in terms of the sensitivity of the homogenized coefficients.
Besides the obviously vanishing derivative $\dlt_\alpha g(\ub)=0$, it holds that 
\begin{equation}\label{eq-opg15}
\begin{split}
\dlt_\alpha\aOm{\ub}{\tilde\vb}\circ\dlt_\alpha\Aop & =
\int_\Om \dlt_\alpha\Aop\eeb{\ub}:\eeb{\tilde\vb}\;,\\
\dlt_\alpha\bOm{p+\bar p}{\tilde\vb}\circ\dlt_\alpha\Bb & =
\int_\Om (p+\bar p) \dlt_\alpha\Bb:\eeb{\tilde\vb}\;,\\
\dlt_\alpha\cOm{p+\bar p}{\tilde q}\circ\dlt_\alpha\Kb & =
\int_\Om \nabla \tilde q \cdot \dlt_\alpha\Kb\nabla( p+\bar p)\;,\\
\dlt_\alpha r(p) &=
\dlt_\alpha \cOm{p+\bar p}{\tilde p}\circ\dlt_\alpha\Kb \\
 & = \int_\Om \nabla \tilde p \cdot \dlt_\alpha\Kb\nabla( p+\bar p)\;.
\end{split}
\end{equation}
Using the ``total pressure'' $P:= p+\bar p$, the following tensors are employed to evaluate the expression in \cref{eq-opg15}:
\begin{equation}\label{eq-opg16}
\begin{split}
  \eeb{\ub}\otimes\eeb{\tilde\vthetabf}\;&, \quad P \eeb{\tilde\vthetabf} \;,\\
  \nabla P\otimes\nabla\tilde q_1 \;&,\quad \nabla P\otimes\nabla\tilde q_2\;, \\
  \nabla \tilde p\otimes\nabla P\;.
\end{split}
\end{equation}
Now, using  these tensors, \cref{eq-opg13} is computed, as follows:
\begin{equation}\label{eq-opg17}
\begin{split}
  & \dlt_\alpha^\tot \Lcal =   - \Lambda_\Psi  \dlt_\alpha r(p) \\
  & + \Lambda_\Phi\left(\dlt_\alpha\aOm{\ub}{\tilde\vthetabf}-
\dlt_\alpha\bOm{P}{\tilde\vthetabf}  +\dlt_\alpha\cOm{P}{\tilde q_1}\right) \\
& + \Lambda_\Psi\dlt_\alpha\cOm{P}{\tilde q_2} + \Lambda_\Xi\pd_\Hop \Xi(\Hop)\dlt_\alpha\Hop
\;.
\end{split}
\end{equation}
Hence the variations of $\Lcal$ \wrt $\Aop,\Bb$ and $\Kb$ are given by the following formulae
\begin{equation}\label{eq-deriv-tensors}
\begin{split}
\dlt_\Aop^\tot \Lcal  & = \Lambda_\Phi \int_\Om  \dlt\Aop_e:\eeb{\ub}\otimes\eeb{\tilde\vthetabf}
\;, \\
\dlt_\Bb^\tot \Lcal  & = -\Lambda_\Phi \int_\Om  \dlt\Bb_e : P\eeb{\tilde\vthetabf}\;,\\
\dlt_\Kb^\tot \Lcal  & = \int_\Om \dlt\Kb_e :
\left(
\Lambda_\Phi \nabla P\otimes\nabla\tilde q_1  \right.\\
&\quad\quad\quad\quad\quad\left. +\Lambda_\Psi \left(\nabla P\otimes\nabla\tilde q_2 - \nabla \tilde p\otimes\nabla P\right)
\right)
\end{split}
\end{equation}
As $\Xi(\Hop)$ solely depends on the regularization parameter $\Rb$, see \cref{eq:filter_function}, we get
\begin{equation*}
\pd_\Hop \Xi(\Hop)\dlt_\alpha\Hop = \int_\Om(\Rb - \FF (\Rb)\cdot(\dlt \Rb - \pd_\Rb\FF (\Rb)\circ\dlt \Rb )    
\end{equation*}
for the regularization term in \cref{eq-opg17}.
In the context of the finite element discretization introduced in \cref{sec:sgp},
the homogenized coefficients are supplied as constants in each element $\Om_e$ of the partitioned domain $\Om$. Accordingly,
the expressions in \cref{eq-opg16} are supplied elementwise at the Gauss integration points.

\subsection{Design parametrization}\label{sec:design_params}
The design of the cell $Y$, that is the decomposition into the solid skeleton $Y_m$ and the pores $Y_c$, can be parameterized in a number of ways. In \cite{Huebner-Solid-2019}, we employed a so-called spline-box structure parameterized by design variables defining positions of the spline control polyhedron. This kind of parametrization is convenient due to its generality to handle quite arbitrary design, but leads to complicated formulations of design constraints which are needed to preserve essential geometrical requirements (\eg positivity of channel crosssections).

\begin{figure}[htb]
	\centering
	\includegraphics[width=0.55\textwidth]{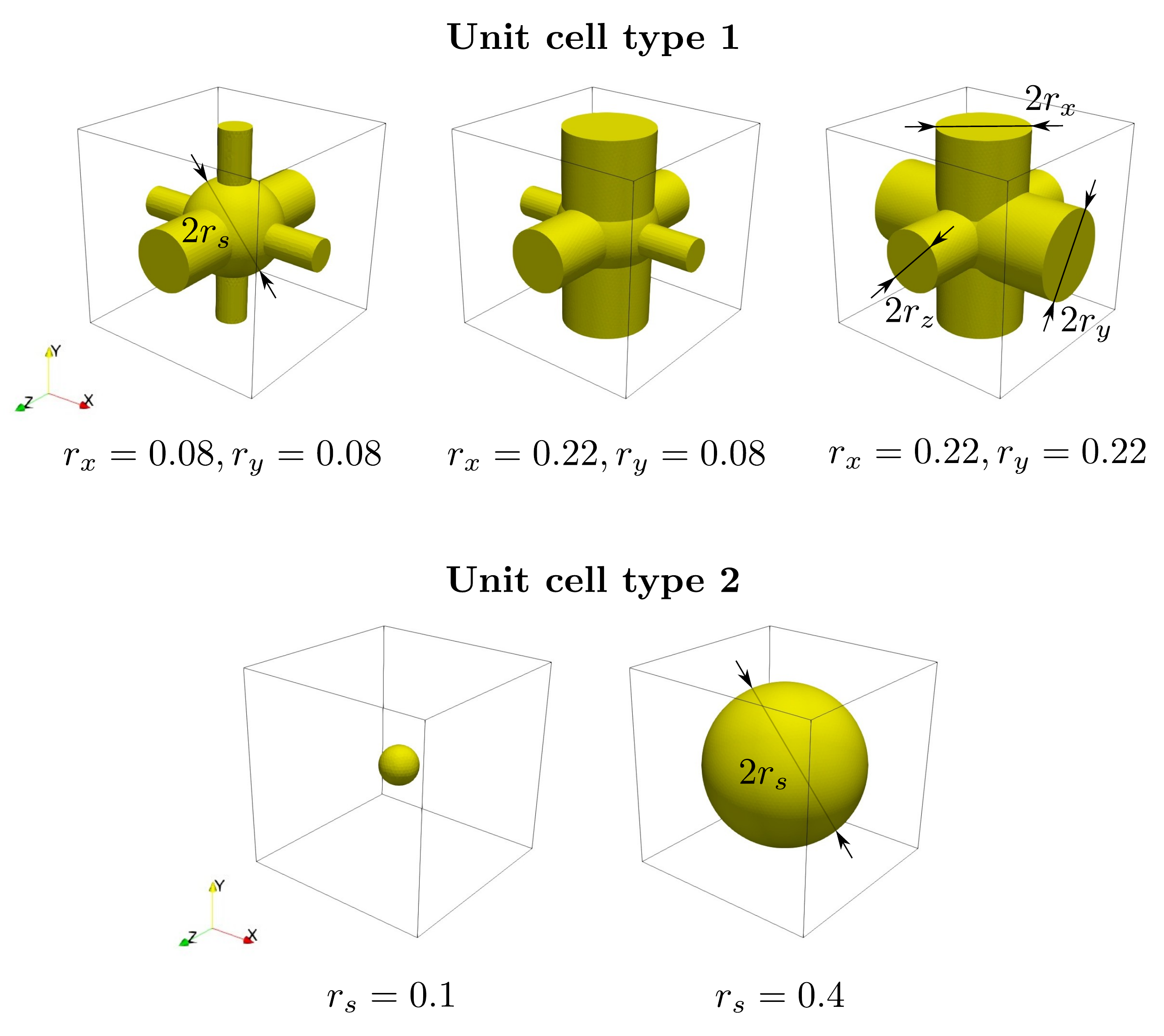}
	\caption{Parametrization of unit cells: unit cell type 1 is parameterized by radii $r_x$ and $r_y$, both ranging from 0.08 to 0.22, $r_z = 0.15$ and $r_s = 0.25$ are kept constant; unit cell type 2 is parameterized by radius $r_s$ ranging from 0.1 to 0.4.}
	\label{fig:design_params}
\end{figure}

In this paper, we employ two specific types of microstructures illustrated in \cref{fig:design_params}, where the channels are shaped as a 3D cross (type 1), or a sphere (type 2). Hence, the latter microstructure is featured by zero permeability and therefore, we consider dry pores (voids) in the mechanical model.
Due to these specific geometries, we can use a rather simple parametrization, which is listed in \cref{tab-alpha}. For a unit cell of type 1, $r_x$ and $r_y$ refer to the radii of the cylinders pointing in $x$- and $y$-direction respectively. The third parameter $\varphi$ describes the cell rotation, about axis $z$. For the unit cell type 2, the spherical voids, whose radii are described by $r_s$, provide an orthotropic material with nearly isotropic elastic properties. Therefore, rotations are not enabled for this cell type. Importantly, box constraints can be imposed on $r_x,r_y$ and $r_s$ straightforwardly to guarantee geometric feasibility.

\begin{table}[h]
\begin{tabular}{l|lll}
  microstructure \# & \multicolumn{3}{l}{cell parameters} \\
  \hline
  1 & $r_x$ & $r_y$ & $\varphi$ \\
  2 & $r_s$ & - & - \\
  \hline
\end{tabular}
\caption{The parametrization of the pore geometry for the two types of the microstructures: 1: the 3D cross, 2: the sphere.}\label{tab-alpha}
\end{table}

\begin{figure}[htb]
	\centering
	\includegraphics[width=0.4\textwidth]{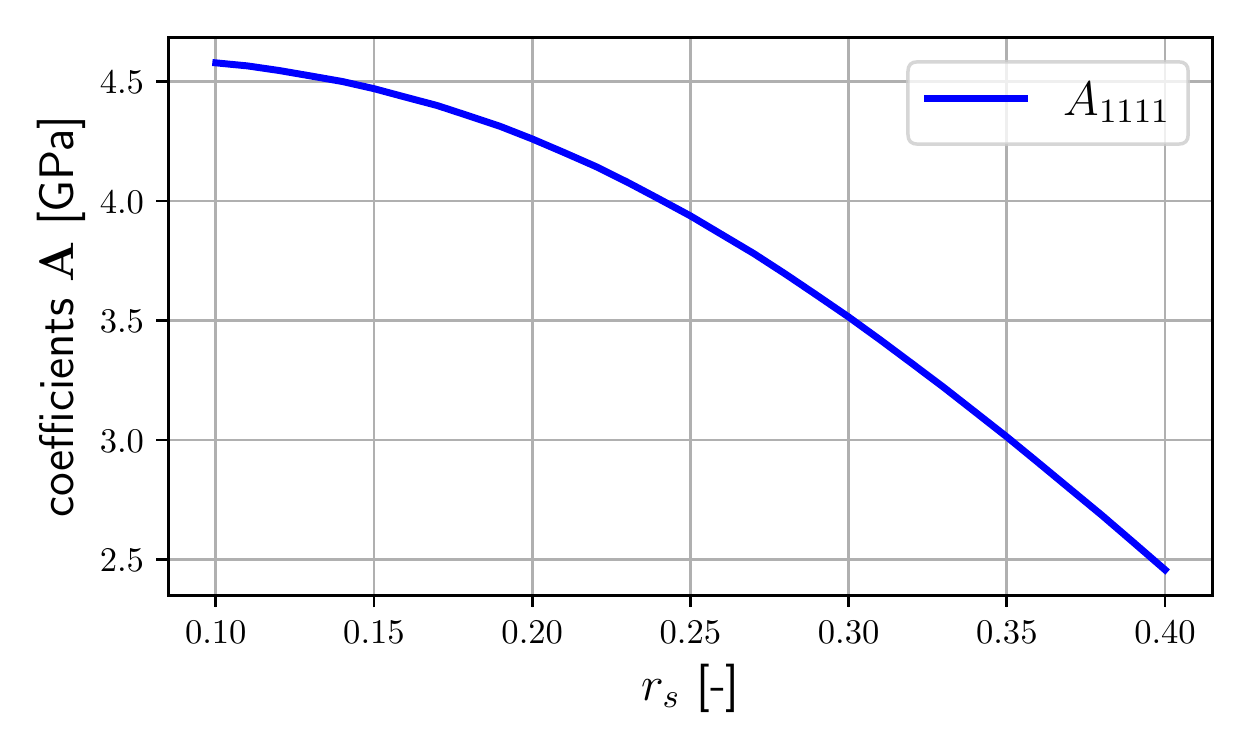}	\caption{Unit cell type 2: dependence of $A_{1111}$ on parameter $r_s$.}
	\label{fig:design_uc2_A}
\end{figure}

\begin{figure}[htb]
	\centering
	\includegraphics[width=0.4\textwidth]{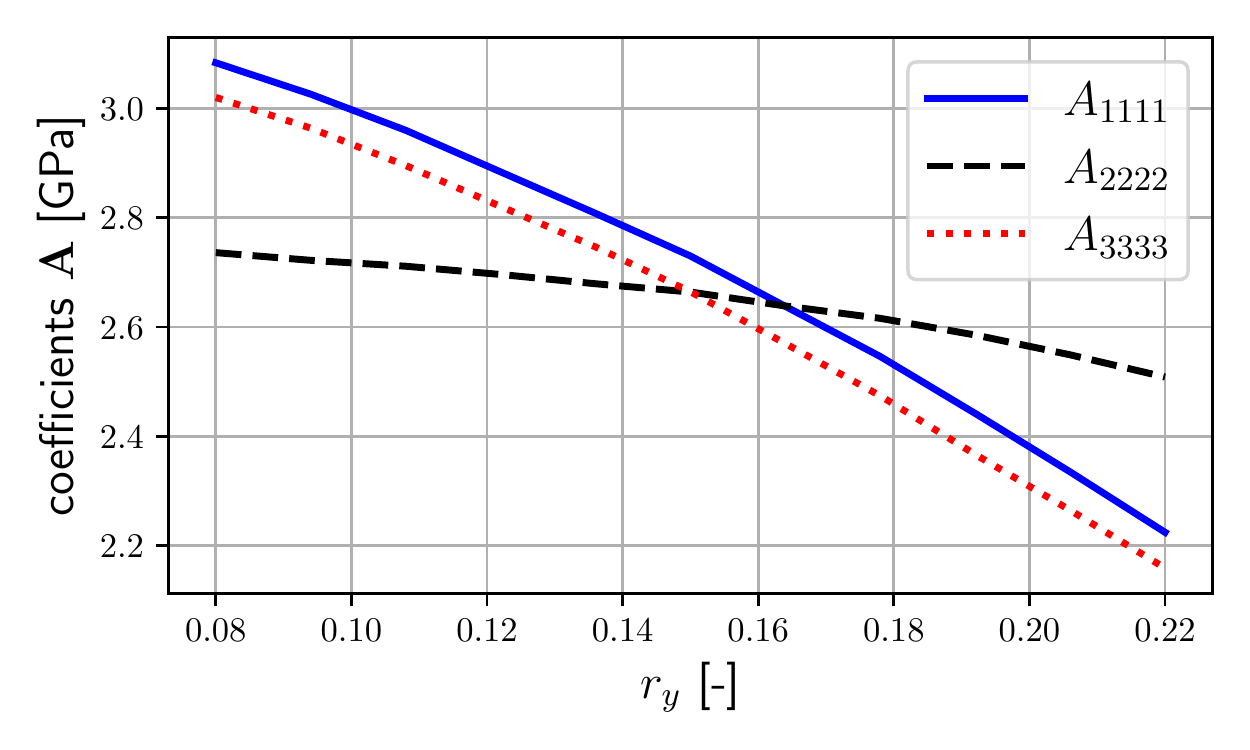}\hfil
	\includegraphics[width=0.4\textwidth]{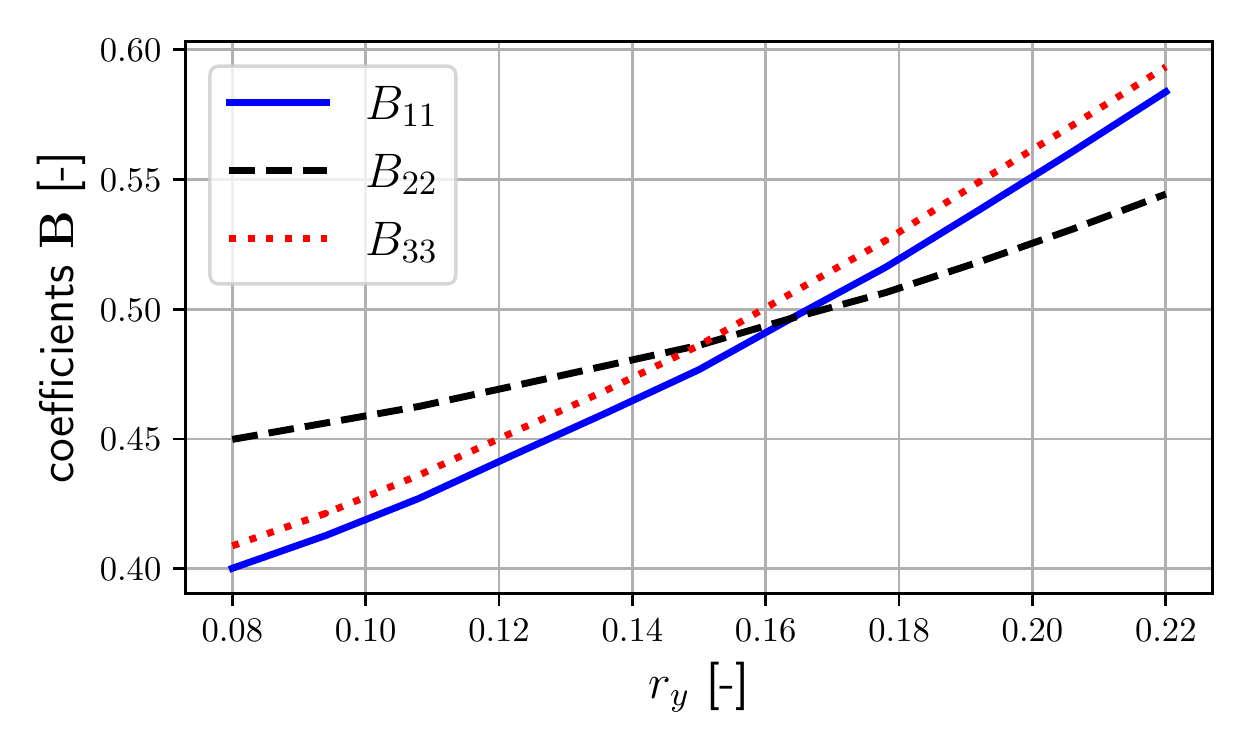}
	\includegraphics[width=0.4\textwidth]{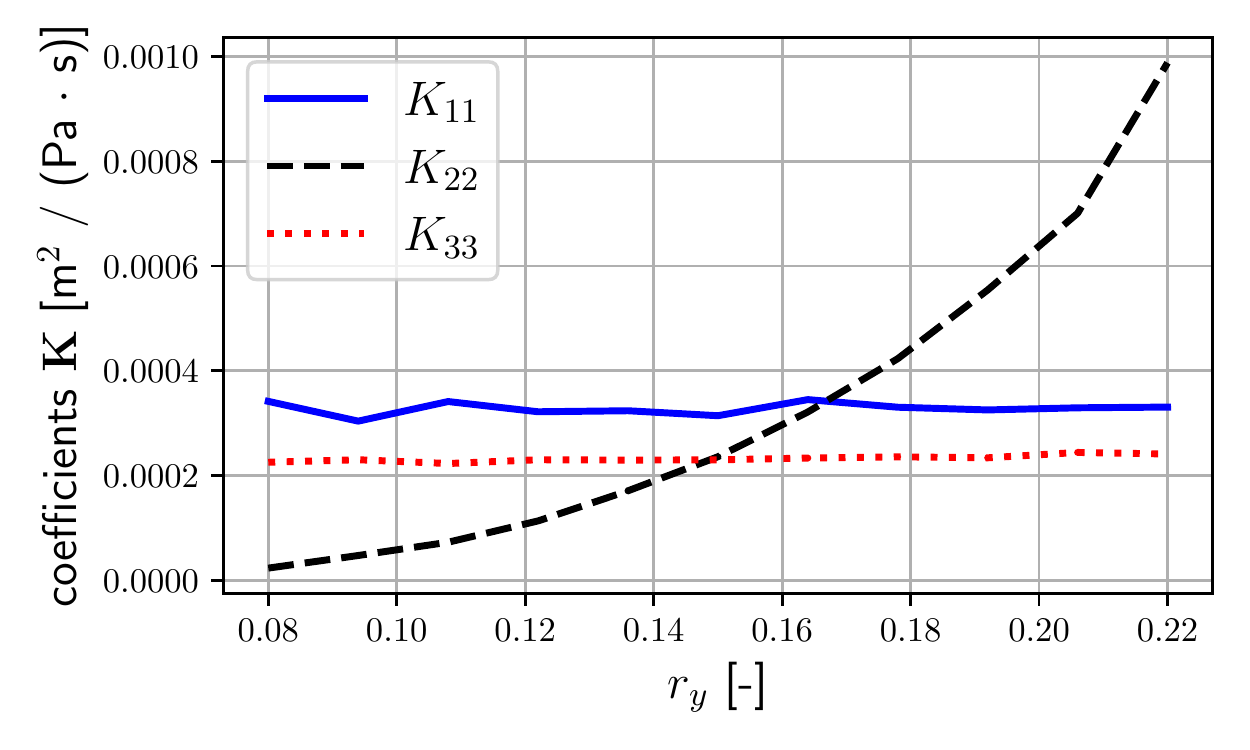}\\
	\caption{Unit cell type 1: dependence of homogenized coefficients $\Aop$, $\Bb$, and $\Kb$ on $r_y$; $r_x = 0.15$ is fixed.}
	\label{fig:design_uc1_ABK}
\end{figure}

To illustrate a sensitivity of the material properties determined by the homogenized coefficients $\Hop$, In \cref{fig:design_uc2_A}, for unit cell type 2, the elasticity as the only relevant material property is displayed as function of $r_s$. In \cref{fig:design_uc1_ABK}, for unit cell type 1, selected components of the poroelastic tensors and of the permeability are reported as functions of $r_y$.
\section{A Sequential Global Programming formulation} \label{sec:sgp}
The basic description of the Sequential Global Programming algorithm along with convergence aspects were presented in \cite{Semmler-SIAM-2018}, where SGP was applied to a multi-material optimization based on a two-dimensional time harmonic Helmholtz state equation. 
The setting and procedure described in this manuscript differs from the one in \cite{Semmler-SIAM-2018} in the following major points: first, in \cite{Semmler-SIAM-2018} a selection of finitely many fixed materials was considered as admissible set. In this paper, each admissible material is computed by homogenizing unit cell, which itself is configurable by a number of geometric parameters. Thus, the designer can choose in each point of the design domain from $M$ different unit cell types \emph{and} adjust the geometric parameters for the latter. 
Second, the SGP approach is extended to a multi-physics setting using a slightly different separable approximation and third, a different solution strategy is employed for the subproblems arising from this. This strategy does not impose any assumption on the parametrization. In particular, parametrizations can be non-analytical and non-differentiable. This leads to a greater design flexibility.  
Despite these differences, there is also an important feature, the approach presented here has in common with the one outlined in \cite{Semmler-SIAM-2018}: separable models are established in terms of (effective) material tensors $\Hop$ rather than their parameterization $\alpha$. Then, the parametrization is directly treated at the level of sub-problems without further 
convexification. Thanks to the separable character of the chosen first order model the resulting generally non-convex sub-problems can - in principal - still be solved to global optimality. 

The advantages of this approach are twofold: first, due to the separable model functions being able to capture also non-convex features of the original cost function typically a low number of outer iterations, equivalently to the number of state problems to be solved, is required; and second, due to the good fit of the separable models with the cost function as well as the fact that non-convex sub-problems are solved to global optimality the overall algorithm is less start value dependent and less prone to be trapped in poor local minima.
This is in contrast to traditional approaches, where a local model is established directly based on the sensitivity of cost functions with respect to the design parameterization $\alpha$.

In the following we first derive a fullly discretized counterpart for a slightly generalized of problem \cref{eq-opH1}.
Then we describe in detail how the separable first order approximations can be constructed and finally present a practical outline of the full SGP algorithm including a generic sub-solver allowing to compute near globally optimal solutions for sub-problems using a brute-force strategy.

\subsection{A fully discretized 2-scale design problem}\label{sec-twoscale-discr}

For the sake of simplicity, the definitions  of sets and functions were introduced in \cref{sec:state-problem,sec:two-scale-opt-problem} based on the assumption that there is only one type of unit cell such that $M=1$. 
Here, for a more general setting, we consider $M$ unit cell types,
each one with $n_i$ design parameters, and introduce index set
$I := \{1, \dots, M\}. $
For each unit cell type $i \in I$, the admissibility set is defined in terms of box constraints and other purely geometrical constraints. By choosing a suitable parameterization, we can identify these with (geometric) parameter sets 
\begin{equation}
    A_i = [\underline{\ab}_i,\overline{\ab}_i] \subset \RR^{n_i}, 
    \label{eq:parameter-sets}
\end{equation}
with $\underline{\ab}_i, \overline{\ab}_i \in \RR^{n_i}$ being lower and upper bound vectors constraining the corresponding parameter vector $\balpha_i \in \RR^{n_i}$. 

\begin{remark}
We note that, while in this manuscript the parameters in \cref{eq:parameter-sets} are always used to vary the geometrical properties of the unit cell, variations in the material parameters could be described in the same way. Thus, SGP can handle both of these situations.
\end{remark}

\begin{figure*}[ht]
    \centering
    \includegraphics[width=0.65\textwidth]{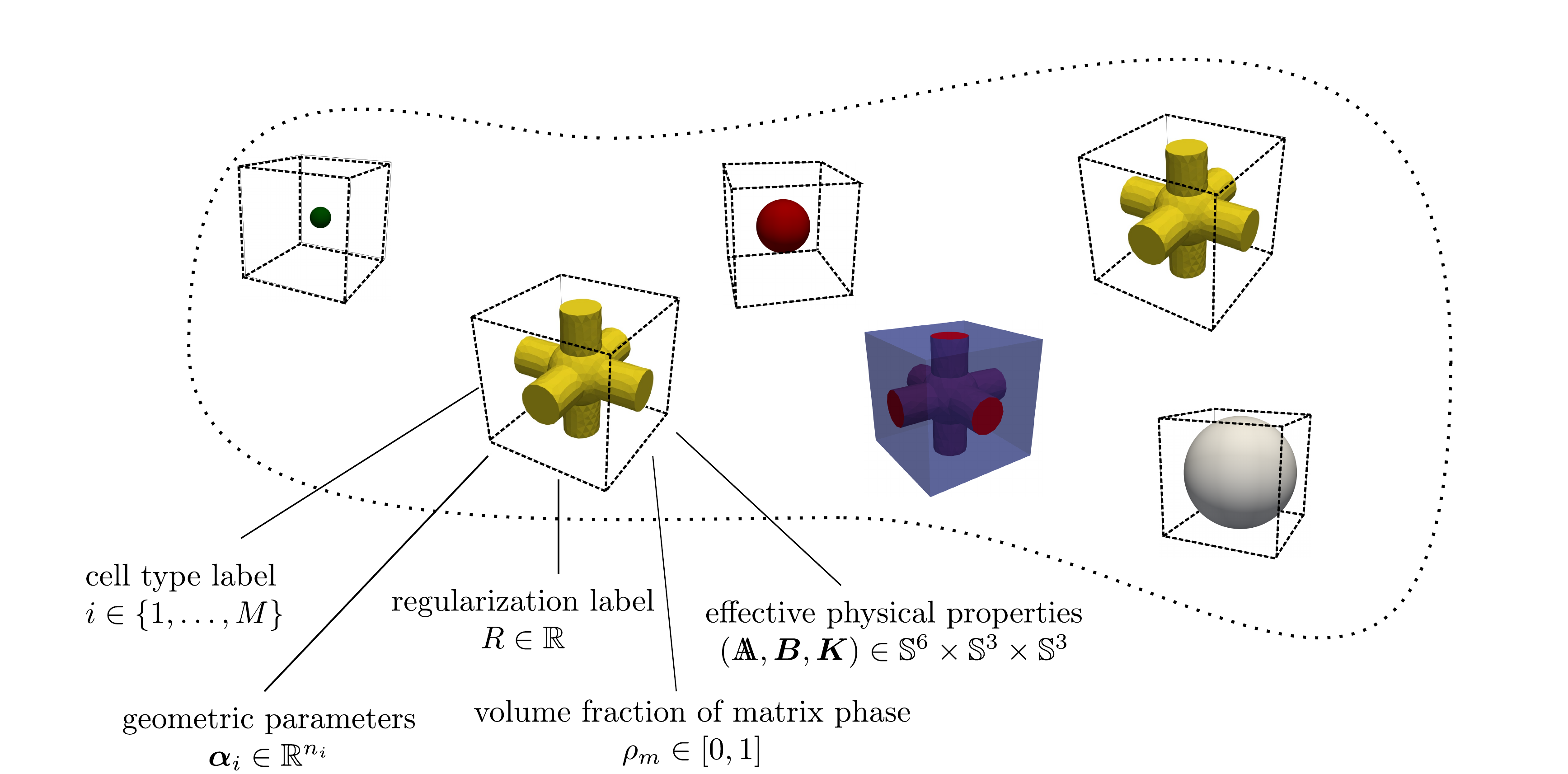}
    \caption{Collection of materials: each material, represented by a unit cell object, comes along with a collection of data such as geometric parameters, physical properties and further labels.}
    \label{fig:material_catalogue}
\end{figure*}

We further define for all $i \in I$ map
\begin{equation}
    \Hcal_i: 
    \begin{cases}
        A_i & \to \Tcal \\ 
        \balpha_i &\mapsto (\Aop, \Bb, \Kb, \rho_m, R),
    \end{cases}
    \label{eq:map-H-continuous}
\end{equation}
where $\Hcal_i(\balpha)$ performs the homogenization procedure described in  \cref{sec:two-scale-opt-problem}. \cref{fig:material_catalogue}
illustrates the components of $\Hcal_i(\balpha_i)$.

We denote the union of the ranges of all $\Hcal_i$ by
\begin{equation}\label{eq:def-union-hom-image}
    H \coloneqq \bigcup_{i=1}^{M} \Hcal_i(A_i)
\end{equation}
and with that generalize the set of admissible design functions to become
\begin{align*} \Uad = & \left\{\Hop \in L^\infty(\Omega;\Tcal) \,\vert\, \Hop(x) \in H \, \right.\\
& \left. \text{ for a.e. } x \in \Omega \right\}.
\end{align*}

Now the state problem operator
\begin{equation}
    \Zcal: 
    \begin{cases}
        \Uad & \to \RR^3 \times \RR \\
        \Hop &\mapsto \zb = (\ub,p),
    \end{cases}
\end{equation}
with displacement function $\ub(\Hop)$ and hydraulic pressure function $p(\Hop)$ reads exactly as before. 

We finally use a slightly more general resource function than in \cref{sec:state-problem,sec:two-scale-opt-problem} as follows:
\begin{equation}\label{eq:vol-func}
    \rho: 
    \begin{cases}
        \Uad &\to \RR \\
        \Hop &\mapsto \rho.
    \end{cases}
\end{equation}
A concretization could be the total volume fraction of a specific material phase (see description of $\rhobarm$ in \cref{sec:two-scale-opt-problem}).

Based on these definitions, we then formulate an FMO-type problem 
\begin{equation}\label{eq-opHgen}
\begin{aligned}
\min_{\Hop \in \Uad} \quad \Fcal(\Hop,\zb) := & \Lambda_\Phi \Phi(\Hop,\zb) + \Lambda_\Psi \Psi(\Hop,\zb)\\
& + \LambXi \Xi(\Hop) \\
\textrm{s.t.} \quad\quad\quad\quad  \zb =&\Zcal(\Hop), \\
  \rho(\Hop) \leq& \rhobarm, 
\end{aligned}
\end{equation}
where $\rhobarm \in \RR$ is the resource constraint value and cost functions and $\Phi$, $\Psi$, $\Xi$ and their weights $\Lambda_\Psi, \Lambda_\Phi, \Lambda_\Xi$ have been already introduced in \cref{sec:two-scale-opt-problem}). \\

Although problem \cref{eq-opHgen} is formulated directly in the tensor variable $\Hop$, a realization of the feasibility condition $\Hop \in \Uad$ would force us to evaluate the homogenization maps $\Hcal_i \; (i\in I)$.
This has the consequence that for each evaluation of the cost function, a homogenization procedure, which contains a series of cell problems, has to be conducted. To alleviate this situation, we follow \cite{BendsoeKikuchi}  and carry out the homogenization procedure only for discrete samples of the design parameter space.
For each unit cell type $i$, we introduce a grid with nodes $A_i ^{\text{nodes}} \subseteq A_i$ and effective material coefficients are only computed, via homogenization, at the sampled nodes of this grid. In addition, we define a piecewise cubic Hermite interpolator for these samples to realize the continuous mapping 
\begin{equation}\label{eq:param-to-tensor}
    \Hcaltilde_i:
    \begin{cases}
        A_i \to \Tcal \\
        \balpha_i \mapsto (\Aop, \Bb, \Kb, \rho_m, R),
    \end{cases}
\end{equation}
for all $i \in I$. 
We denominate this procedure as the \emph{offline} phase of a two-scale optimization approach, as it can be performed independent from the \emph{online} optimization procedure that is subject to constraints, that go beyond the box constraints on the parameter sets as in \cref{eq:parameter-sets}.

\begin{figure*}[ht]
    \centering
    \includegraphics[width=0.7\textwidth]{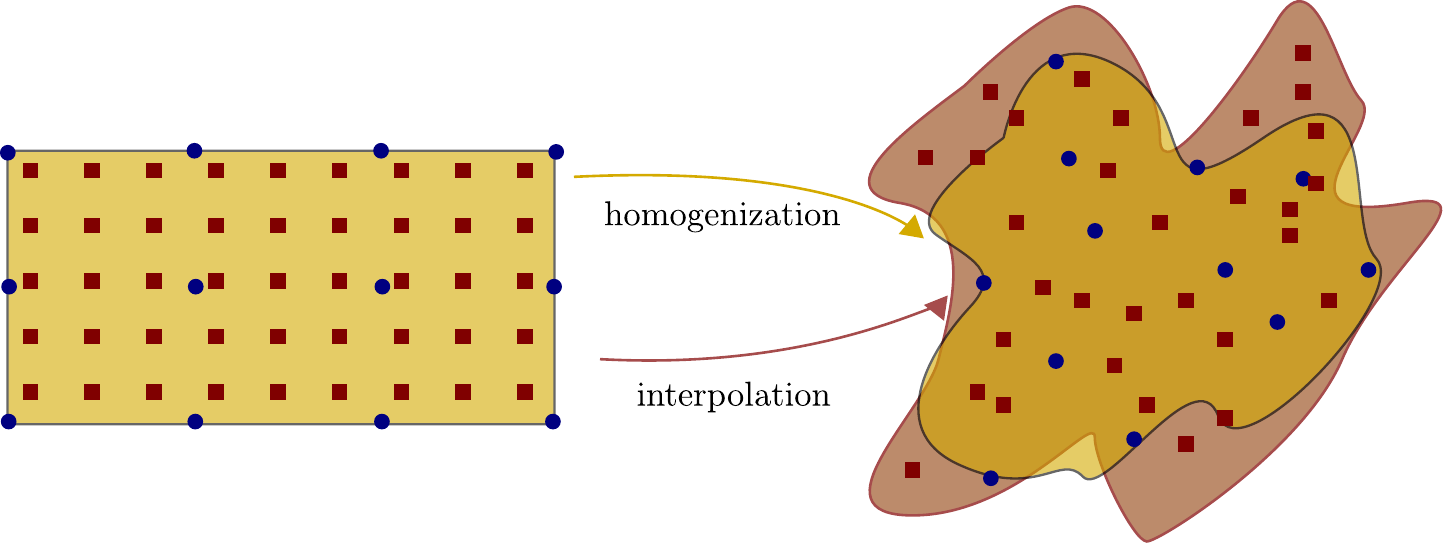}
    \caption{Left: Sketch of parameter set $A_i$ and samples from its subsets $A_i^{\text{nodes}}$ (blue dots), that serves as a construction basis of interpolated $\Hcaltilde_i$, and $\Agrid_i$ (red squares), on which the optimization process is performed. In general, $A_i^{\text{nodes}}$ and $\Agrid_i$ can be fully independent from each other. Right: Simplified sketch of the original effective material coefficients spaces $\Hcal_i(A_i)$ (yellow surface) and the the images of interpolated $\Hcaltilde_i(A_i)$ (red surface). The blue dots and red squares represent the images of the parameters from respectively $A_i^{\text{nodes}}$ or $\Agrid_i$.}
    \label{fig:sketch-images-H}
\end{figure*}

For the case $M=1$, the conventional approach would be now, to perform the optimization based on the interpolated functions $\Hcaltilde_1$ over the full parameter set $A_1$. This is not directly possible for $M>1$. One way to get around this would be to introduce another interpolation between the different unit cell types similar as it is done in discrete material optimization (DMO) \cite{hvejsel2011}. Rather than that we introduce design grids
\begin{equation}
\Agrid_i \subset A_i, \; i \in I,
\end{equation}
for all unit cell types. Only elements of $\Agrid_i, \; i \in I$ will be considered in the optimization process later. This way, in general, only an approximate solution of the design problem can be computed. However it will turn out that this strategy combines well with the separable non-convex model introduced later in \cref{sec:subproblems}. Moreover the resulting error can be easily controlled by the distance and number of samples in $\Agrid_i, \; i \in I$. 
The relation of different grids and mappings for the material coefficients are visualized and elaborated in \cref{fig:sketch-images-H}.\\

As we only optimize on $\Agrid_i, \; i \in I$, \cref{eq:def-union-hom-image} is approximated by
\begin{equation}\label{eq-Htilte}
    \Htilde \coloneqq \bigcup_{i=1}^{M} \Hcaltilde_i(\Agrid_i).
\end{equation}
We note that elements of $\Htilde$ can be precomputed already in the \emph{offline} phase. In general, this leads to a higher memory requirement, but additionally reduces \emph{online} computation time.

Finally, we briefly introduce a finite element approximation, with $\nel$ finite elements, and therefore introduce element index set
$
E \coloneqq \{1, \dots, \nel\}
$  
to indicate a finite element distinctively by its index $e \in E$.
We further assume that the design is constant on each element and can thus be represented by $$ \Hopd \in \Htilde^{\nel}
$$ 
We remark that through the definition of $\Htilde$ in \cref{eq-Htilte} this condition already states that only material tensors are eligible, for which a unit cell type $i$ and a parameter vector $\balpha_i$ in $\Agrid_i$ exists.
Moreover, we replace physical functions $\Phi$ and $\Psi$, regularization function $\Xi$ and solution operator $\Zcal$ by their discretized counterparts, \eg
\begin{equation}
    \Zcal_h:
    \begin{cases}
        \Htilde^{\nel} \to \RR^{\ndof} \\
        \Hopd \mapsto (\ubm,\pbm)
    \end{cases},
\end{equation}
where $\ndof$ is the dimension of the discrete state solution space.
The discretized version of resource function $\rho$ \cref{eq:vol-func} is 
\begin{equation}\label{eq:vol-func-h}
    \rho_h:
    \begin{cases}
        \Htilde^{\nel} \to \RR \\
        \Hopd \mapsto \rho_h.
    \end{cases}
\end{equation}

The optimization problem, fully discretized in design and state space, then reads
\begin{equation}
\begin{aligned}\label{eq:sgp-opt-problem}
\min_{\Hopd \in \Htilde^{\nel}}  \max_{\lambrho \in \RR^+}\quad & \Fcal_h(\Hopd,\zbm,\lambrho) \\
\textrm{s.t.} \quad & \zbm =\Zcal_h(\Hopd),
\end{aligned}
\end{equation}
with
\begin{equation*}
\begin{aligned}
\Fcal_h(\Hopd,\zbm, \lambrho) := & \Lambda_\Phi \Phi_h(\Hopd,\zbm) + \Lambda_\Psi \Psi_h(\Hopd,\zbm)\\
& +\lambrho \left(\rho_h(\Hopd)-\rhobarm\right) + \LambXi \Xi_h(\Hopd).
\end{aligned}
\end{equation*}

We note that we have eliminated the resource constraint by the Lagrange formalism. Later we will suggest to use a bisection strategy as introduced in \cite{sigmund99} for the framework of the well known OCM method.
We finally specialize the regularization term to become
\begin{equation}\label{eq:filter_function}
    \Xi_h(\Hopd) = \frac{1}{2}\|\Rb- \FF (\Rb) \|^2,
\end{equation}
where $\FF$ denotes a standard density filter function (see, \eg \cite{bourdin-filter}) with
\begin{equation}
    \FF: \RR^\nel \to \RR^\nel.
\end{equation}
and $\Rb$ is the vector of regularization labels associated with all finite elements $e \in E$.

\subsection{Construction of subproblems}
\label{sec:subproblems}
For any sequential programming algorithm  first a sequence of subproblems has to be defined. Here, in each iteration $k$, we construct \emph{separable} first order approximations, about an expansion point $\Hopd^k \in \Htilde^\nel$, for the components of cost function 
\begin{equation}
    \Jcal(\Hopd, \lambrho) := \Fcal_h(\Hopd,\zb, \lambrho)
\end{equation}
of the original optimization problem in \cref{eq:sgp-opt-problem}. The model problem is
\begin{align}\label{eq:model-problem}
    \min_{\Hopd} \max_{\lambda_\rho \in \RR} \quad&\Jsep\left(\Hopd, \lambrho; \Hopd^k \right) 
\end{align}
where our model function is defined as
\begin{align}\label{eq:model-function}
& \Jsep\left(\Hopd,\lambrho; \Hopd^k\right) := \sum_{e\in E} \Jsepe\left(\Hopd_e, \lambrho;\Hopd_e^k\right) \nonumber\\
            &= \sum_{e\in E} \Jphystilde\left(\Dopd_e;\Dopd_e^k\right) + \lambrho \Jvoltilde((\rho_m)_e)  \nonumber\\
            & + \LambXi \Jregtilde(R_e;R_e^k) + \Lambg \Jglobtilde(\Aop_e;\Aop_e^k) 
 \end{align}
 with
\begin{align*}
            & \Dopd_e := (\Aop_e,\Bb_e,\Kb_e) \in \cS^6 \times \cS^3 \times \cS^3, \nonumber\\
            & \Dopd_e^k := (\Aop_e^k,\Bb_e^k,\Kb_e^k) \in \cS^6 \times \cS^3 \times \cS^3, \nonumber\\
            &\Hopd, \Hopd^k \in\, \Htilde^\nel.
 \end{align*}
In the following, we describe each component of $\Jsep$ in more details.\\
For this, we split $\Jcal(\Hopd, \lambrho)$ as
\begin{equation*}
    \Jcal(\Hopd, \lambrho) = \Jphys(\Hopd) 
    + \lambrho \Jvol(\Hopd) + \LambXi \Jreg(\Hopd)
\end{equation*}
with 
\begin{align}
    \Jphys(\Hopd) & := \Lambda_\Phi \Phi_h(\Hopd, \zb) + \Lambda_\Psi \Psi_h(\Hopd,\zb), \label{eq:Jphys} \\
    \Jvol(\Hopd) & := \rho_h(\Hopd) - \rhobarm, \label{eq:Jvol}\\
    \Jreg(\Hopd) & := \Xi_h(\Hopd).\label{eq:Jreg}
\end{align}

From tuple $\Hopd$, only the effective material coefficients $\Aop, \Bb$ and $\Kb$, are relevant for $\Jphys$. 
Consequently, for $\Jphys$, we define a separable approximation of type 
\begin{equation}
    \sum_{e \in E} \Jphystilde\left(\Dopd_e;\Dopd_e^k\right),
\end{equation} 
where $\Jphystilde$ is the following generalization of the first-order MMA-like model suggested in \cite{stingl-siam-2009} for functions defined in tensor variables:
\begin{align}\label{eq:Jphys_approx}
&\Jphystilde\left(\Dopd_e;\Dopd^k\right) \nonumber\\
&= C_{\text{phys}} - \left\langle \Aop_e^k \left[\frac{\partial \Jphys(\Dopd^k)}{\partial \Aop}\right]_e\Aop_e^k, \Aop_e^{-1} \right\rangle_{\cS^6} \nonumber \\
&- \left\langle \Bb_e^k \left[\frac{\partial \Jphys(\Dopd^k)}{\partial \Bb}\right]_e \Bb_e^k, \Bb_e^{-1} \right\rangle_{\cS^3} \nonumber \\
&- \left\langle \Kb_e^k \left[\frac{\partial \Jphys(\Dopd^k)}{\partial \Kb}\right]_e \Kb_e^k, \Kb_e^{-1} \right\rangle_{\cS^3}.
\end{align}
Here $C_{\text{phys}}$ is a constant that is chosen to establish the zeroth order correctness of the model and $<\cdot,\cdot>_{\{\cS^6,\cS^3\}}$ denotes the Frobenius inner products for matrices from $\cS^6$ and $\cS^3$, respectively. It is further mentioned that in contrast to the model in \cite{stingl-siam-2009}, we refrain from working with flexible generalized asymptotes $L_e^\Aop \in \cS^6, L_e^\Bb, L_e^\Kb \in \cS^3$, but simply choose all of them to be zero matrices. The partial derivatives of $\Jphys$ with respect to the material coefficients $\Aop,\Bb$ and $\Kb$ can be easily extracted from the expressions in \cref{eq-deriv-tensors}. 

The function $\Jvol$ that describes the fraction of utilized matrix material, is separable by definition, and depends solely on $\rho_m$. We accordingly choose
\begin{equation}\label{eq:Jvol-approx}
    \Jvoltilde((\rho_m)_e;\rho_m^k) = (\rho_m)_e.
\end{equation}

The function $\Jreg$ given in \cref{eq:Jreg} solely depends on the regularization label $\Rb \in \RR^\nel$, which is a component of tuple $\Hopd \in \Htilde^\nel$.
The separable approximation of $\Jreg$ is thus of the form
\begin{equation}
    \sum_{e \in E} \Jregtilde(R_e;\Rb^k),
\end{equation}
where 
\begin{align}\label{eq:Jreg-approx}
    &\Jregtilde(R_e;\Rb^k)  \\
    & = \frac{1}{2}\, \bigg\Vert \Rtilde\left(R_e;\Rb^k\right) - \left[\FF \left(\Rtilde\left(R_e;\Rb^k\right)\right)\right]_e \bigg\Vert^2.\nonumber
\end{align}
In \cref{eq:Jreg-approx}, we further employ function
\begin{equation*}
    \Rtilde\left(R;\Rb^k\right) \coloneqq  \left(R_1^{k}, \dots, R_{e-1}^{k}, R, R_{e+1}^{k}, \dots, R_{\nel}^{k} \right),
\end{equation*}
in which the regularization label is varied only in the $e$-th entry by value $R$, and contributions of expansion point $\Rb^k$ are used in the neighboring entries. Is is noted that \cref{eq:Jreg-approx} can be reduced to a convex quadratic function of type
\begin{equation*}
    a_e R_e^2 + b_e R_e + c_e,
\end{equation*}
by precomputing $a_e,b_e,c_e \in \RR$, which are independent from $R_e$.

Finally, we implement a step size control for the design from one iteration to the next one by adding 
\begin{equation}\label{eq:sgp-glob}
    \sum_{e \in E}\Jglobtilde\left(\Aop_e,\Aop_e^k\right) = \sum_{e \in E}\frac{1}{2}\left\Vert \Aop_e - \Aop_e^k \right\Vert^2
\end{equation}
with a positive factor $\Lambg$ to the model cost function. Alternatively, a more general globalization strategy, similar to the regularization approach with regularization label $R$ in \cref{eq:Jreg-approx}, could be pursued by introducing particular globalization labels. Here, we assume that evaluating the design step size based on the stiffness tensor $\Aop_e$ and $\Aop_e^k$ is sufficient, and, in particular, the uniqueness of the globalization labels, such that
\begin{equation}
    \Aop_e = \Aop_e' \Rightarrow \balpha_e = \balpha_e',
\end{equation}
is satisfied.

\subsection{The SGP algorithm with a brute-force sub-solver}
\label{sec:SGPalg}
Having at hand the separable first-order approximations of the objective function and penalization terms, we are now able to formulate the iterative scheme that is described by \cref{alg:sgp-multimat}. We make extensively use of the separable structure of 
\begin{equation*}
 \Jsep\left(\Hopd, \lambrho; \Hopd^k\right)= \sum_{e \in E} \Jsepe\left(\Hopd_e, \lambrho;\Hopd_e^k\right) 
\end{equation*}
and solve the subproblems, of each iteration $k$, for each finite element $e \in E$ individually. 
This is done by evaluating $\Jsepe$ for all (finitely many) $\Hopd_e \in \Htilde$ and, based on these evaluations, identifying a global minimizer $\Hopd_e^*$. Note that, with each $\Hopd_e$, a unique geometric cell label $\alpha_e$ is associated and thus, by determining $\Hopd_e^*$, we also determine respective $\alpha_e^*$ and material class index $i^*$. As mentioned already earlier a bisection strategy is applied to treat the resource constraint, see \cref{alg:sub-problem} for the details. To keep things simple, it is assumed that the resource constraint is always active at a minimizer. If no resource constraint is applied, the outer loop in \cref{alg:sub-problem} is simply omitted. 

After each iteration, the original cost function $\Jcal$ is evaluated with the current solution of the subproblems $\Hopd_e^*$. If a descent in $\Jcal$ was achieved, we continue the iterative process. If not, we employ the step width control, by increasing multiplier $\Lambg$ of globalization term \cref{eq:sgp-glob}, and resolve the subproblems using \cref{alg:sub-problem}.
\begin{algorithm}[ht]
	\caption{Sequential Global Programming for parametrized multi-material optimization}
	\label{alg:sgp-multimat}
  \begin{algorithmic}[1]
  	\State $k \gets 0$\;
  	\State initialize $\Hopd^0 \in \Htilde^\nel$\;
        \State $\Jdiff \gets \infty$\;
  	\While{$ \Jdiff > 0 \; \text{and}  \; k \leq k_{\rm max} $}
  	    \State initialize $\Lambg \in \RR$\;
            \While{$\Jdiff < 0$}
  			\State $\Hopd_{\Lambg}^* \gets$ solve \cref{eq:model-problem} to global \newline \hspace*{6.5em} optimality using \cref{alg:sub-problem} \;
  			\State increase $\Lambg$\;
            \EndWhile
  		\State $\Hopd^* \gets \Hopd_{\Lambg}^*$\;
            \State $\Jdiff \gets \Jcal(\Hopd^k) - \Jcal(\Hopd^*) $
  		\State $k \gets k +1$  \;    
    \EndWhile
  \end{algorithmic}  
\end{algorithm}

\begin{algorithm}[htb]
	\caption{Solve subproblems via brute force strategy} \label{alg:sub-problem}
  \begin{algorithmic}[1]
    \State initialize $\lambrho \in \RR$ for volume bisection
    \While{volume constraint is not satisfied}
      	\ForAll{finite element $e \in E$}
      		\ForAll{unit cell types $i \in I$} 
      		  \State $\balpha_{i}^* \gets $ minimizer on $A_i^\text{grid}$ \;
      		\EndFor\;
      		\State $\balpha^* \gets$ minimizer among all $\balpha_{i}^* \; (i \in I)$\;
                \State $i^* \gets$ unit cell type index of $\balpha^*$ \;
      		\State $\Hopd_e^* \gets $ evaluate $\Hcaltilde_{i^*}(\balpha^*)$     
                       (see \cref{eq:param-to-tensor})
      	\EndFor
            \State $\rho \gets$ evaluate $\rho_h(\Hopd_e^*)$ (see \cref{eq:vol-func-h});
            \If{$\rho > \rhobarm$}
                \State increase $\lambrho$\;
            \Else
                \State decrease $\lambrho$\;
            \EndIf
    \EndWhile  	
  \end{algorithmic}    
\end{algorithm}

\section{Numerical results}\label{sec:numerical-results}
In this section, we demonstrate the abilities of SGP by means of numerical examples.
It is build up successively by first increasing the design freedom to the two-scale optimization problem, while observing the respective optimized designs and then studying the effect of regularization.

In \cref{sec:one-cell}, we start with the unit cell that is constructed by three intersection fluid channels, visualized in the top row of \cref{fig:design_params}, and study the impact of the micro-structure's local orientation on the performance of the optimized designs. It will be seen that, thanks to the strength of our model, we do neither have to use smart initial orientations, as proposed \eg in \cite{pedersen1989,norris2006} by aligning the anisotropic material \wrt principal directions of the stress tensor, nor we have to enforce artificially a regular design. 

Then, we present a pareto front and investigate the influence of different weightings of compliance and fluid flux, in the cost function, on the resulting designs. When we proceed from one point on the Pareto front to the next one, we intentionally refrain from using the previous design as a warm start. Nevertheless and despite the non-convex character of our weighted cost function, Pareto curves are obtained, in which none of the points is dominated by another one. We trace this observation back to the ability of the SGP method to avoid poor local solutions.

In \cref{sec:two-cells}, we proceed to demonstrate the ability of SGP to handle more than one unit cell type. We again compute a Pareto curve for this case. It will be observed that the new Pareto front is, due to the increase in the design freedom, is strictly dominating the previous one. It will be observed that the more complex parametrization does on average not lead to an increase in the number of state problems to be solved per optimization run.
Note that for the settings presented in \cref{sec:one-cell} and \cref{sec:two-cells}, it was not necessary to employ a globalization strategy to control design changes from one iteration to the next one. Thus, we set the globalization parameter $\Lambg = 0.$ \\

In the end, in \cref{sec:two-cells-reg}, we apply a filtering technique onto the design parameters to both control the speed of variation of local orientation, as well as the interface length between the two unit cell types. Here, we also employ the globalization term described in \cref{eq:sgp-glob}.

\begin{figure}[htb]
	\centering
	\includegraphics[width=0.31\textwidth]{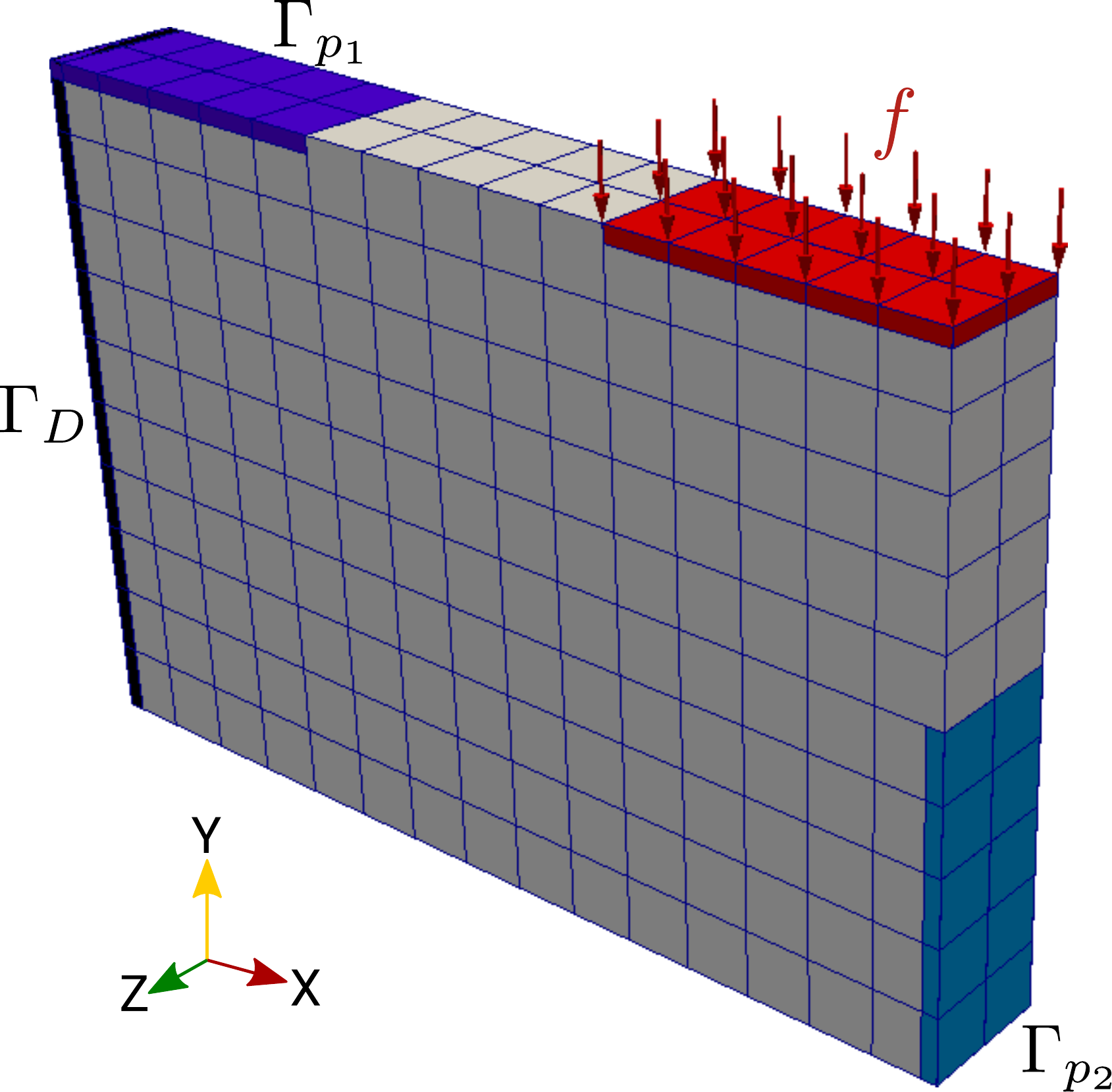}
	\caption{Setup of the macroscopic problem: mechanical traction force $f \!=\! (0,-1,0)^\top$ acts on a part of the body's surface (red) while support is provided on $\Gamma_D$ and pressure values $p_1=1.0$ and $p_2=0.5$ are prescribed on $\Gamma_{p_1}$ and $\Gamma_{p_2}$. The design domain is discretized by 15 x 10 x 2 hexahedra.}
	\label{fig:macro-setting}
\end{figure}

The setting of the poroelastic problem is depicted in \cref{fig:macro-setting}. It is  a recapitulation of the macroscopic problem setting from \cite{Huebner-Solid-2019}, where the authors selected a finite element from the macroscopic domain and optimized the shape of the local microstructure via a spline box approach. In the present paper we provide an extension to this example by solving the two-scale optimization problem with the SGP method described in \cref{sec:sgp}. We note that we work with a rather coarse discretization of the macroscopic domain. The reason is that such a discretization is sufficient to demonstrate the capabilities of SGP as described above. On the other hand, it is readily seen in \cref{alg:sub-problem} that the number of macroscopic elements enters the computational complexity for SGP linearly. Thus, in principle there is no obstacle to work with finer discretizations. 

\subsection{Optimization with one unit cell type}\label{sec:one-cell}
\begin{figure}[h]
    \centering
    \includegraphics[width=0.24\textwidth]{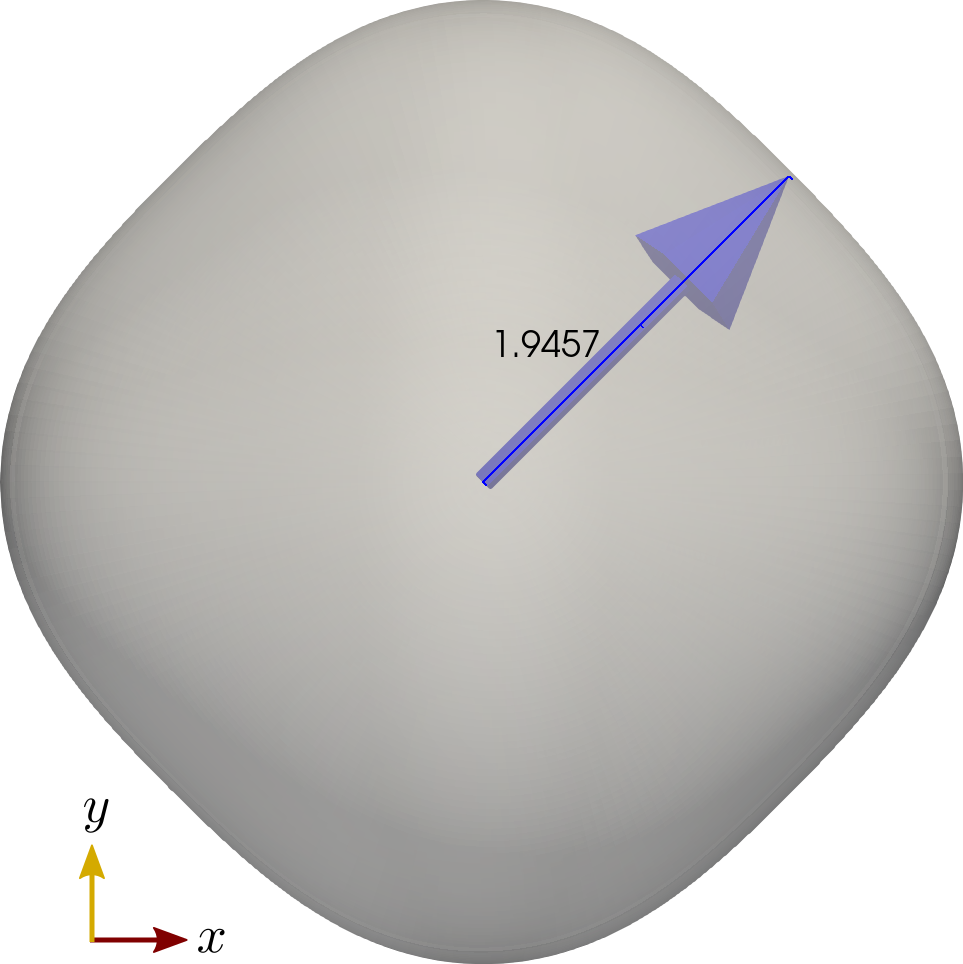}
    \caption{Visualization of directional stiffness of unit cell with maximally opened fluid channels ($r_x = 0.22, r_y = 0.22$). This spherical plot was generated by drawing the entry $A_{1111}$ of the rotated material tensor $\Aop \in \cS^{6}$ for varying rotation angles $(\theta, \phi) \in [0,2\pi]^2$ about $z$- and $y$-axes. For instance, the sketched arrow points to $(\pi/2,0)$ and its length of 1.9457 comes from first entry of the material tensor that is rotated by $\pi/2$ about the $z$-axis.}
    \label{fig:matviz_alpha-022}
\end{figure}

\begin{figure}[htb]
    \centering
    \includegraphics[width=0.27\textwidth]{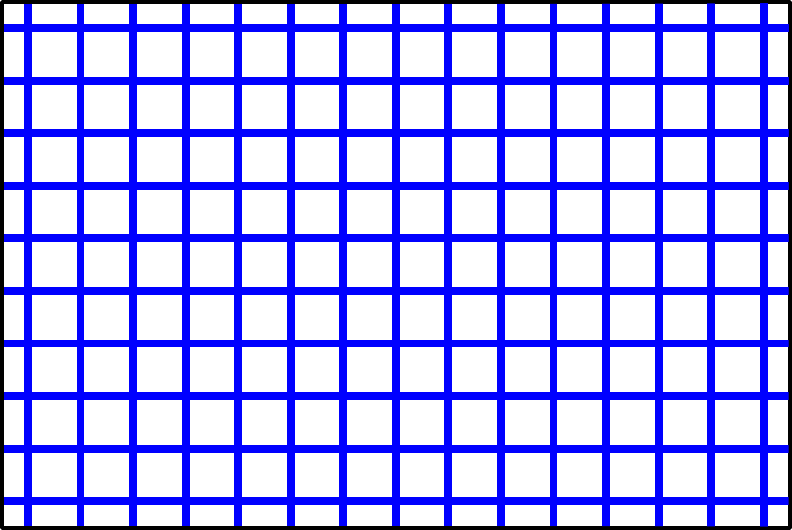}
    \caption{Homogeneous initial design with ${r_x=r_y=0.15}$ and no cell rotation and physical performance $\Phi_\text{init} = 28.9$ and $\Psi_\text{init} = 0.135$.}\label{fig:init-015-015}
\end{figure}

In this section, we employ unit cell type 1, depicted in \cref{fig:design_params}. The geometry consists of three joint cylindrical fluid channels, filled with Glycerine (Young's modulus \SI{4.35}{\giga\pascal}, dynamic viscosity \SI{0.95}{\pascal\second}), that are perpendicular to each other and intersect a hollow sphere in the middle of the cell domain. These channels are embedded in matrix material made of Polystyrene with Young's modulus of \SI{3.9}{\giga\pascal} and dynamic viscosity of \SI{0.34}{\pascal\second}. The feasible range for the geometric design parameters is $A_1 = [0.08,0.22]^2$. Thus, in each finite element $e \in E$, we have the design parameters $\balpha_1 = (r_x,r_y)^\top \in A_1$ to steer the radii of the channels pointing in $y$- and $x$-direction. The radius of the fluid channel that points in $z$-direction (out-of-plane) is kept constant. At the boundaries of the design parameter space, the volume fractions of the stiff material phase are $\rho\left(\Hcal_1\left([0.08,0.08]^\top\right)\right) = 0.7154$ and $\rho\left(\Hcal_1\left([0.22,0.22]^\top\right)\right) = 0.879$. The directional stiffness of the softest version of this unit cell is visualized in \cref{fig:matviz_alpha-022} by means of a polar plot.
The interpolation of $\Hcal_1$ is based on $A_1^{\text{nodes}}$. Here, $A_1^{\text{nodes}}$ is the parameter grid spanned by the components of $\balpha_1$, and for each component we chose 11 equally spaced samples. The subproblems of the SGP algorithm are solved based on the discrete parameter grid $A_1^\text{grid}$. For this grid, we chose a sample size of 28 for each of the two channel radii; again the samples are equally spaced.\\
For the following optimization results with the weighted sum formulation of structural compliance and fluid flux, we employ an initial design guess, visualized in \cref{fig:init-015-015}, that is neither particularly favorable for the mechanical nor for the fluid flow state.

\begin{figure}[htbp]
    \centering
    \subfloat[Optimized design ($z=0$)\label{fig:lamb-10-no-rot-z0}]{\includegraphics[width=0.22\textwidth]{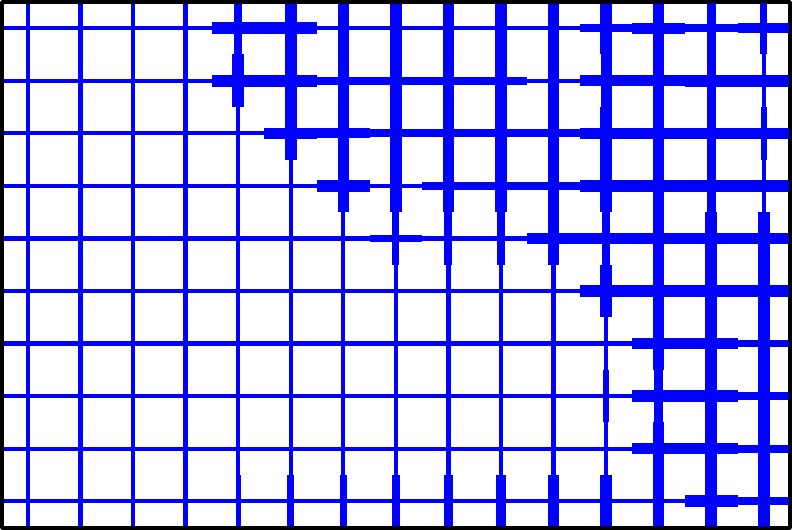}}\quad
    \subfloat[Optimized design ($z=1$)\label{fig:lamb-10-no-rot-z1}]{\includegraphics[width=0.22\textwidth]{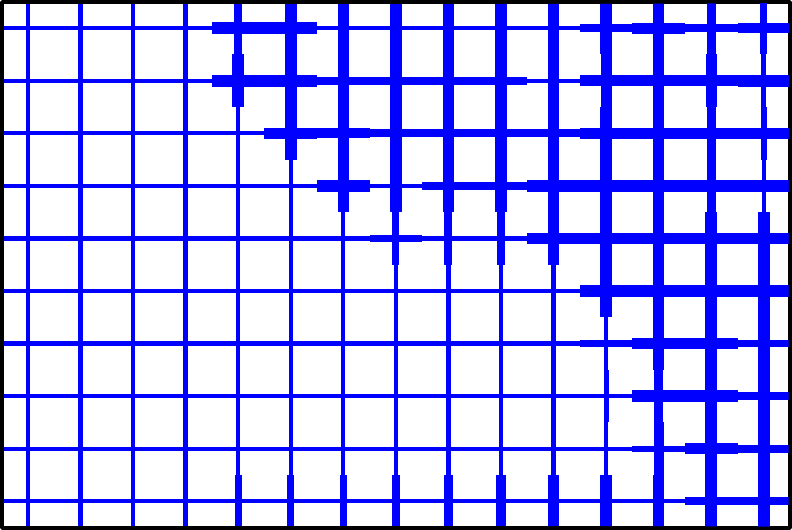}} \\
    \subfloat[Mechanical state\label{fig:lamb-10-no-rot-mech}]{\includegraphics[width=0.24\textwidth]{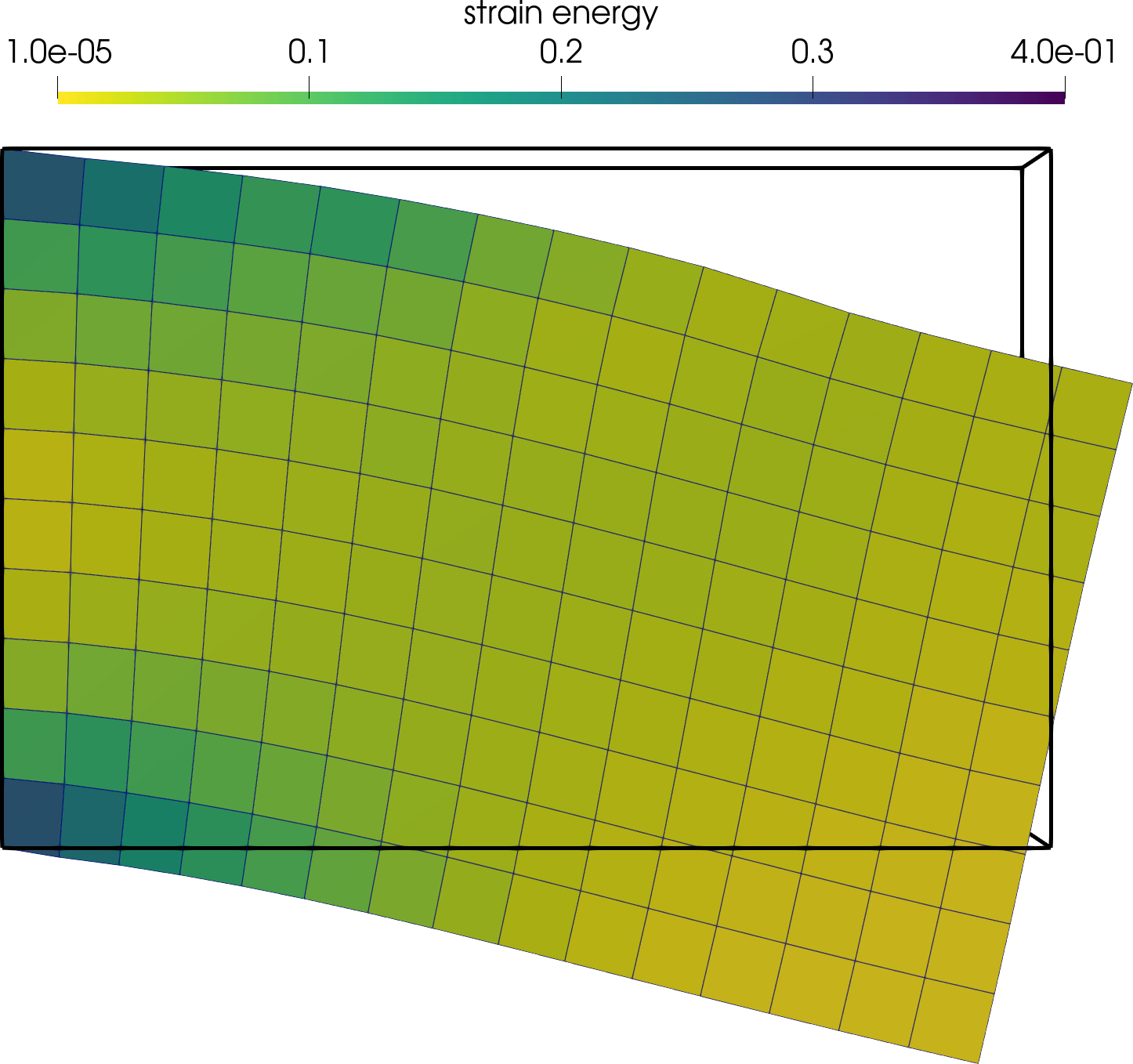}} \\
    \subfloat[Pressure field]{\includegraphics[width=0.24\textwidth]{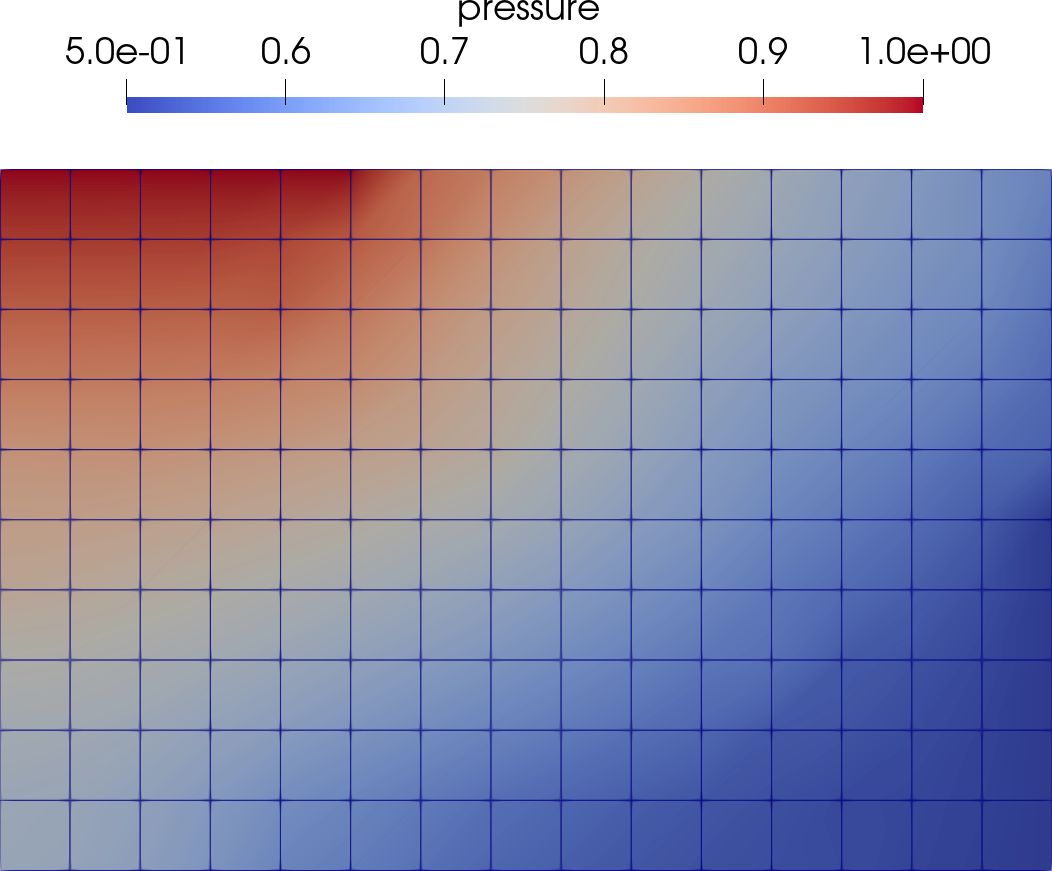}} \\
    \subfloat[Velocity field\label{fig:lamb-10-no-rot-velo}]{\includegraphics[width=0.24\textwidth]{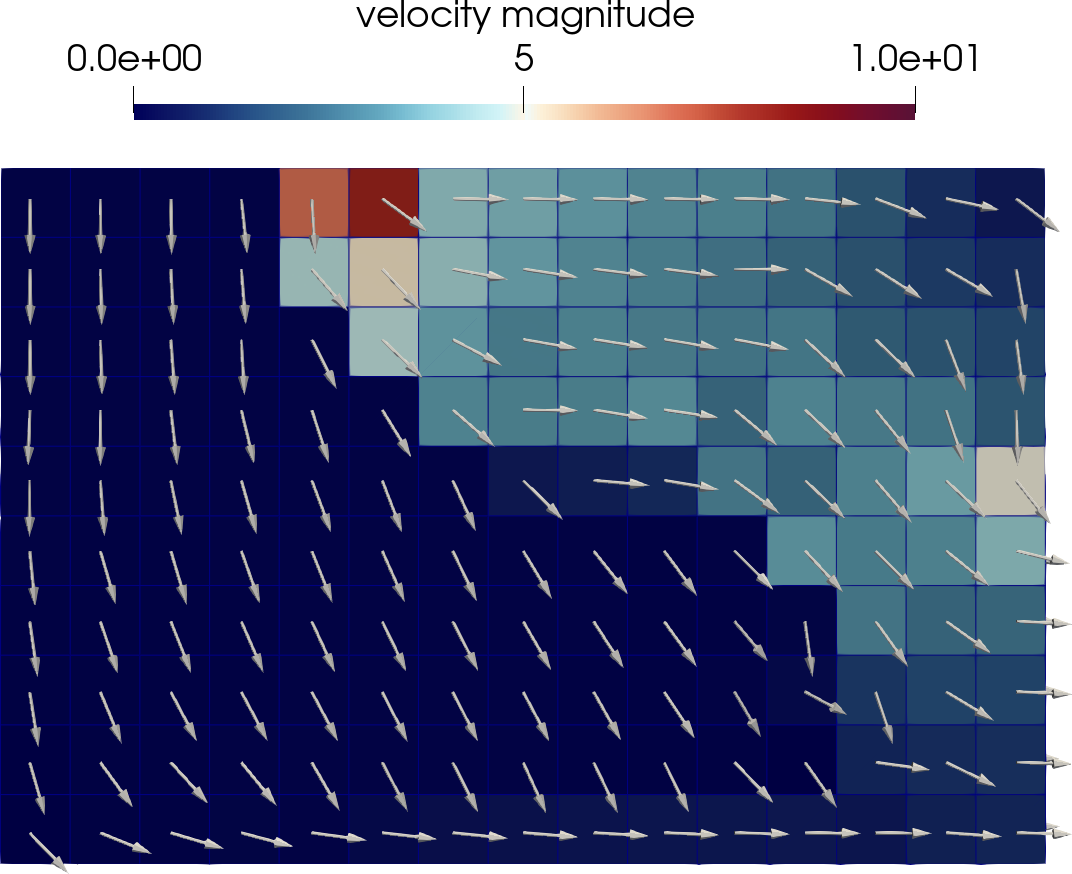}}
    \caption{Optimization result for $\Lambda_\Psi= -10$ and fixed local micro-structure orientation (no rotation) with $\Phi_\text{opt}=27.25$ and $\Psi_\text{opt}=0.275$ for the optimized design in \protect\subref{fig:lamb-10-no-rot-z0},\protect\subref{fig:lamb-10-no-rot-z1}. The initial guess is the design shown in \cref{fig:init-015-015}. In \protect\subref{fig:lamb-10-no-rot-mech} the mechanical state of the optimized design is visualized by deforming the domain by the physical displacements. The strain energy is shown in colors. In \protect\subref{fig:lamb-10-no-rot-velo}, the flow direction is visualized by equally scaled arrows and the colors indicate magnitude of the flow field.}
    \label{fig:min-poroel_no-rot_lambda-10}
\end{figure}

For the described setting, we choose $\Lambda_\Psi = -10$ and obtain the optimized design shown in \cref{fig:lamb-10-no-rot-z0}. Note that the design domain is discretized by two finite element layers in $z$-direction. We made the experience that, for all numerical results presented in this paper, the differences of optimized designs at layer $z=0$ and layer $z=1$ are so small such that they cannot be visually discernible. For this reason, we will only show optimized designs for layer $z=0$ in the rest of the paper.

\begin{figure}
    \centering
    \subfloat[\label{fig:comp-Agrid1-10}]{\includegraphics[width=0.22\textwidth]{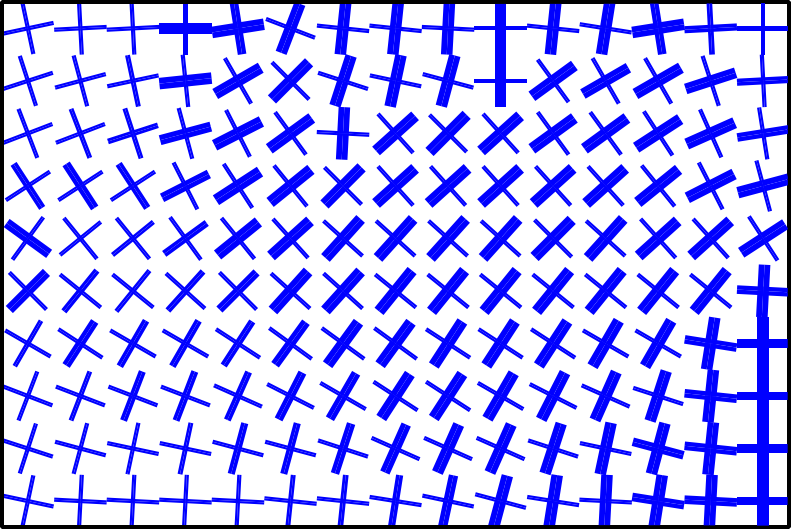}} \quad
    \subfloat[\label{fig:comp-Agrid1-28}]{\includegraphics[width=0.22\textwidth]{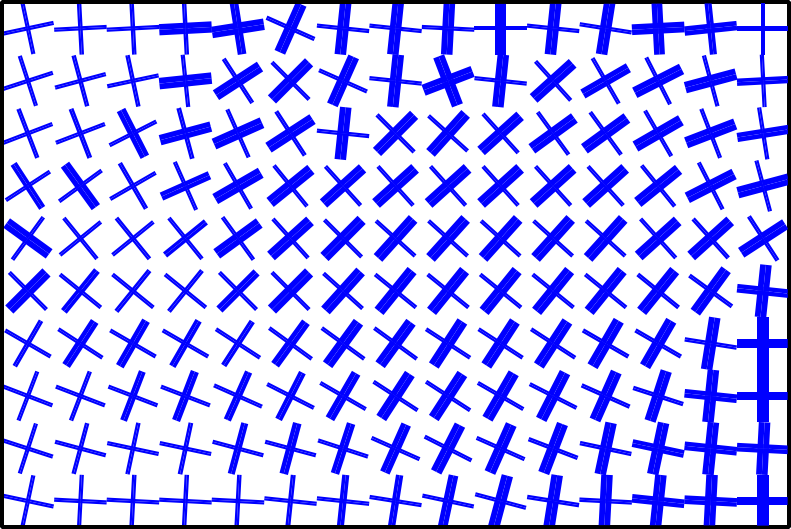}}
    \caption{Two optimized designs for different sample sizes of $\Agrid_1$. \protect\subref{fig:comp-Agrid1-10} 10 samples each for $r_x$ and $r_y$ and 180 samples for $\varphi$. \protect\subref{fig:comp-Agrid1-28} 28 samples each for $r_x$ and $r_y$ and 180 samples for $\varphi$. Here, $\Lambda_\Psi=1$ and $\Lambda_\Psi=-10$. The visual differences are barely perceptible, although \protect\subref{fig:comp-Agrid1-28} has a 1.5\% lower compliance and a 1.7\% higher flux than \protect\subref{fig:comp-Agrid1-10}.}
    \label{fig:comp-Agrid1}
\end{figure}

SGP stopped after 19 iterations, because the difference between the objective values of the old and new design was found to be 0. We note that this comparably low number of iterations is related to the fineness of the design discretization. Thus, using more grid points could lead to a slightly larger number of iterations. On the other hand, in those experiments that we performed in this direction, the visualizations of the obtained result could be hardly distinguished, see \cref{fig:comp-Agrid1}. This is why we do not report results for different choices of $\Agrid_i, \; i \in I$. 
A second observation we can make is that the fluid channels in resulting designs are fully connected. This is due to the fact that no rotational design degrees of freedom were used. On the other hand we will see next that the performance is getting way better, if also local rotations of the micro-structures are allowed.

\subsubsection{Optimized local in-plane rotation of micro-structure}
We introduce angle variable $\varphi \in [0,\pi]$ to allow in-plane rotation, about the $z$-axis, of the micro-structure. The effective material coefficients are rotated by $\varphi$ with the following analytical expressions:
\begin{align}
    \Aop_\text{rot}(r_x,r_y,\varphi) &= \Qb_6(\varphi) \Aop(r_x,r_y)\Qb_6(\varphi)^T, \nonumber\\
    \Bb_\text{rot}(r_x,r_y,\varphi) &= \Qb_3(\varphi) \Bb(r_x,r_y)\Qb_3(\varphi)^T, \nonumber\\
    \Kb_\text{rot}(r_x,r_y,\varphi) &= \Qb_3(\varphi) \Kb(r_x,r_y)\Qb_3(\varphi)^T, 
\end{align}
where $\Qb_6 \in \RR^{6 \times 6}$ are rotation matrices for the stiffness tensor $\Aop$ in Voigt notation and $\Qb_3 \in \RR^{3 \times 3}$ are rotation matrices for the Biot coupling and permeability tensor. We note that no additional evaluation of the homogenization operators are required, as, instead of the micro-structure, the effective material tensors are rotated.  $\varphi$ is discretized with $180$ steps for the brute force approach to solve the SGP subproblem with \cref{alg:sub-problem}. 

\begin{figure}[ht]
    \centering
    \subfloat[Design after one iteration\label{fig:example-poroel-lambdaPsi-10-iter-1}]{
    \includegraphics[width=0.22\textwidth]{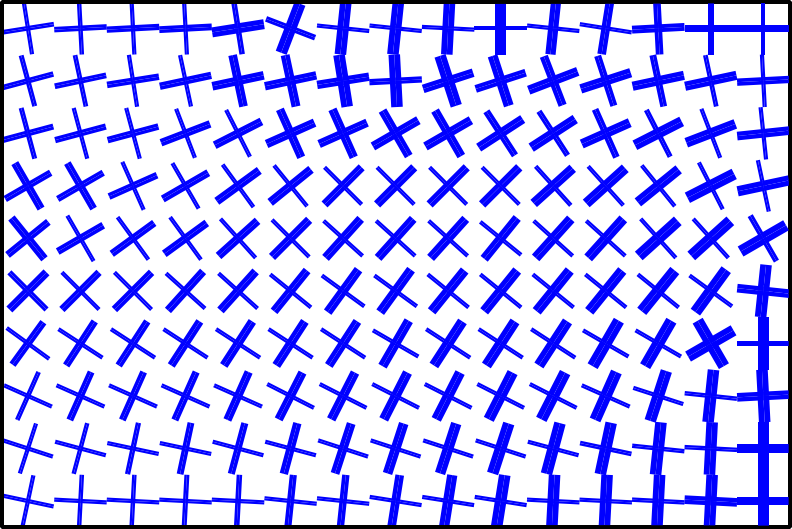}} \,
    \subfloat[Optimized design\label{fig:example-poroel-lambdaPsi-10-optDesign}]{ \includegraphics[width=0.22\textwidth]{poroel_alpha28_phi180_beta60_15x10x2_init015_phi0_lambda10_optDesign}} \\
    \subfloat[Pressure field]{\includegraphics[width=0.24\textwidth]{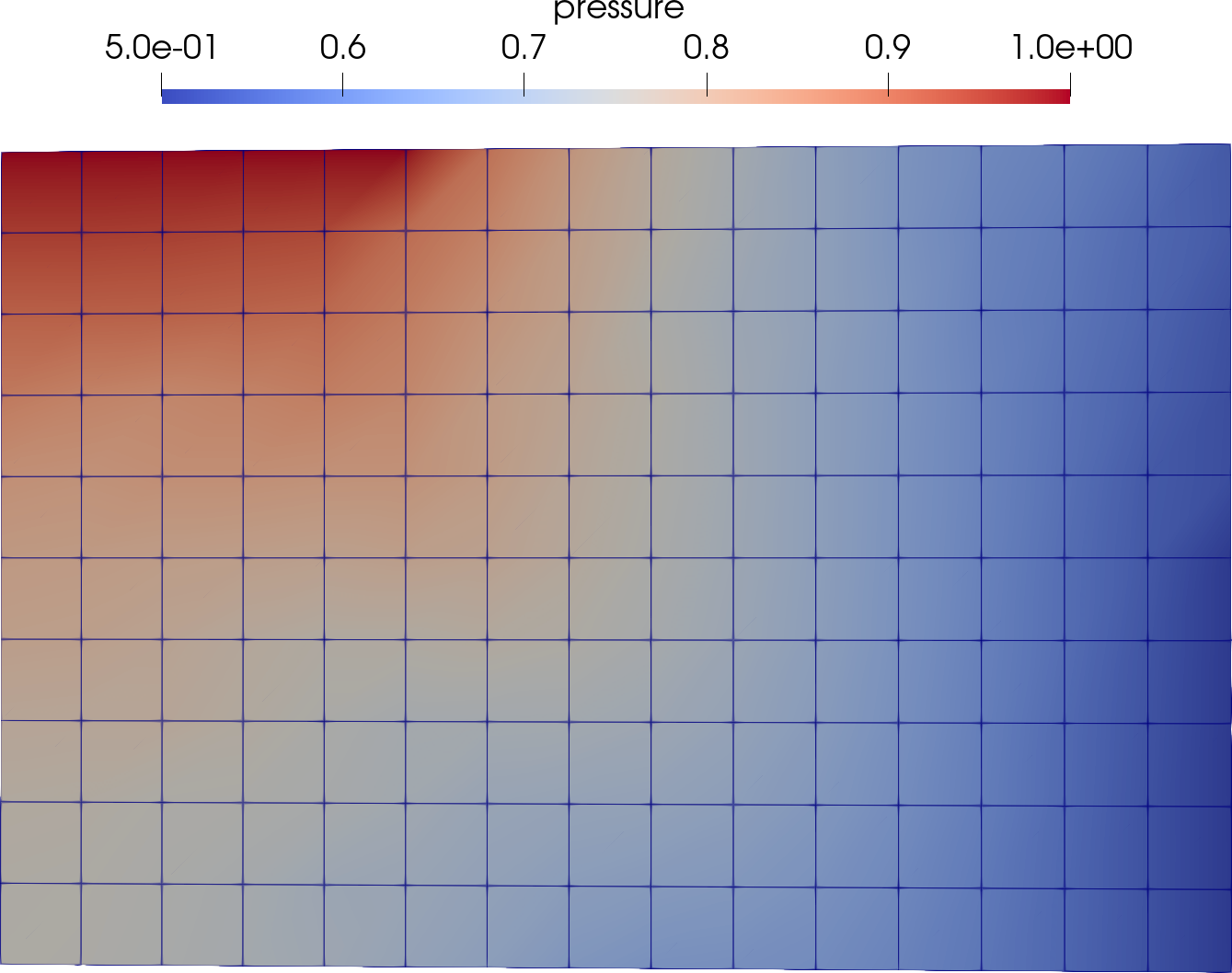}} \quad
    \subfloat[Velocity field]{\includegraphics[width=0.24\textwidth]{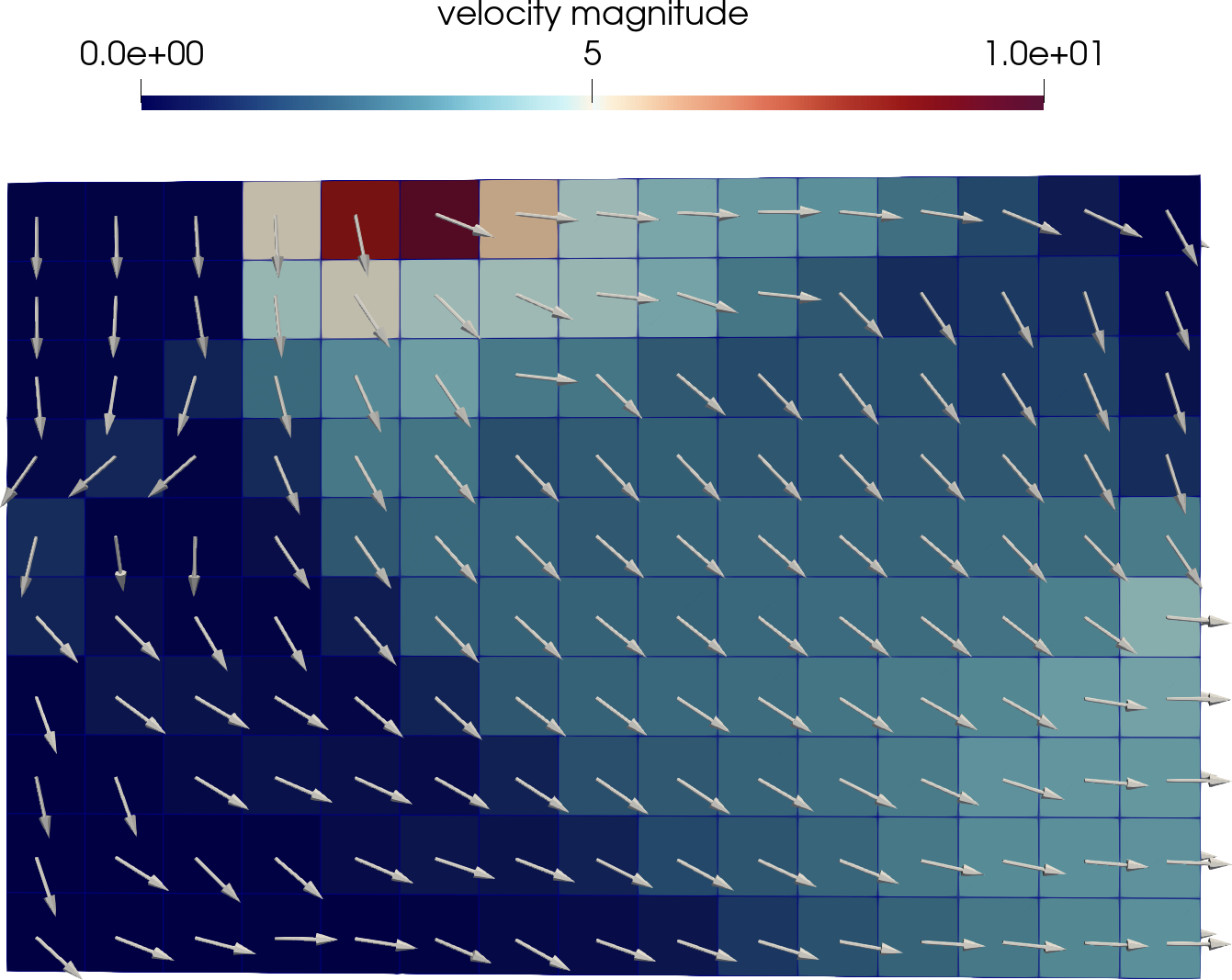}} \quad
    \subfloat[Mechanical strain]{\includegraphics[width=0.24\textwidth]{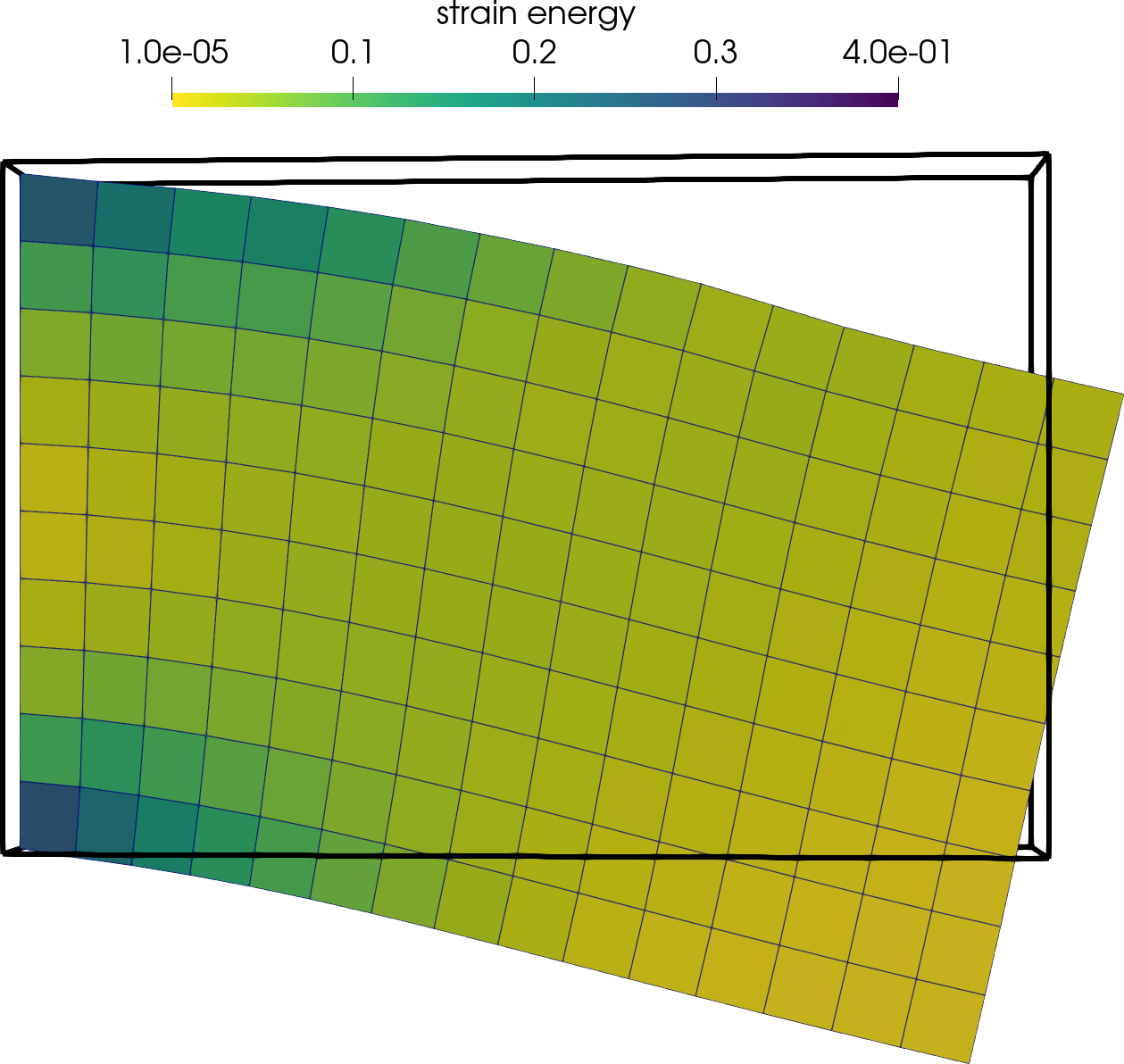}} \\
    \caption{Optimized design with rotational design degrees of freedom and respective physical state for $\Lambda_\Phi = 1$ and $\Lambda_\Psi = -10$, with $\Phi_\text{opt}=27.1$ and $\Psi_\text{opt}=0.413$.}
    \label{fig:poroel_rot_lambda-10}
\end{figure}

\begin{figure}
    \centering
    \includegraphics[width=0.29\textwidth]{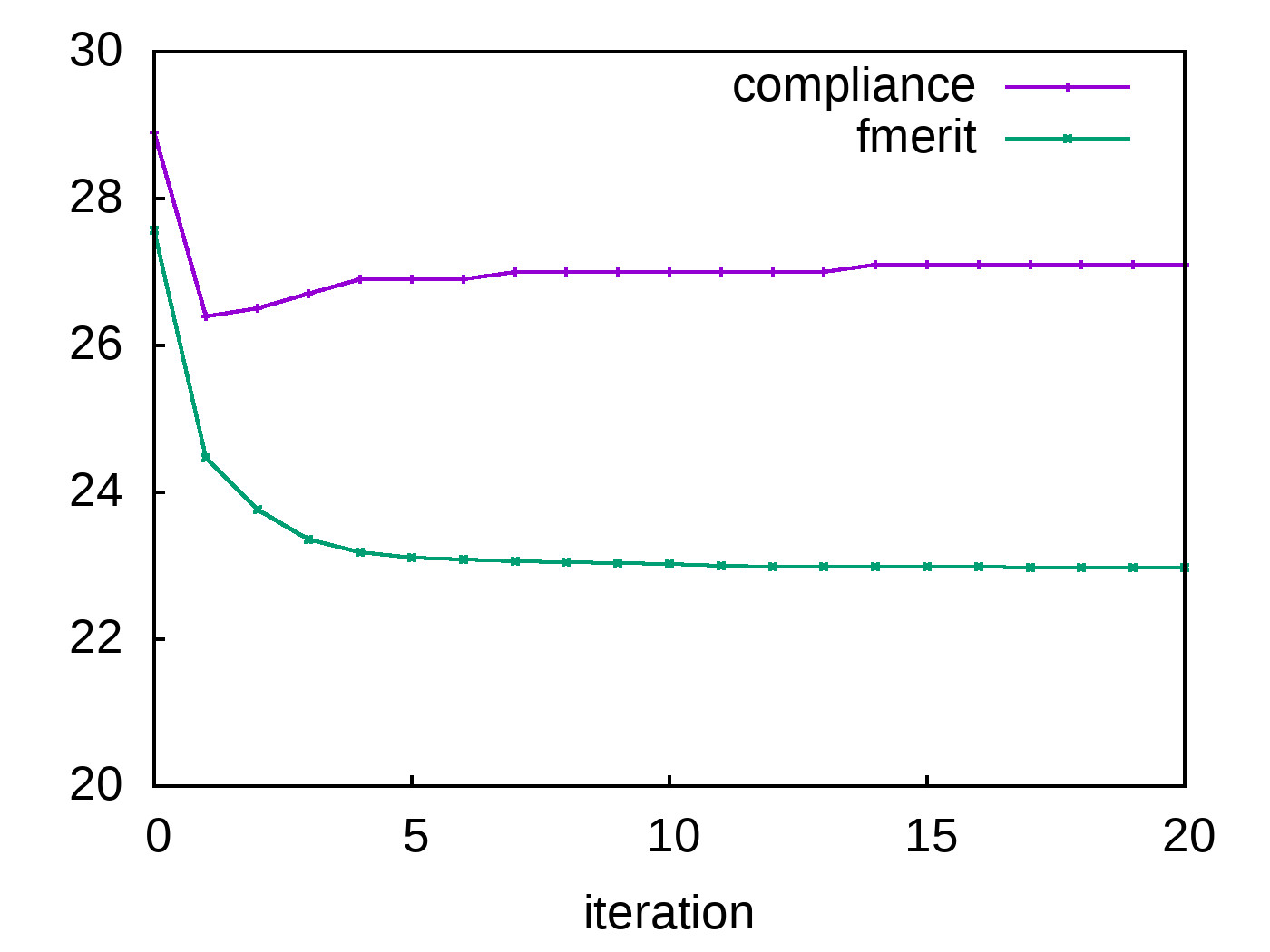} \quad
    \includegraphics[width=0.36\textwidth]{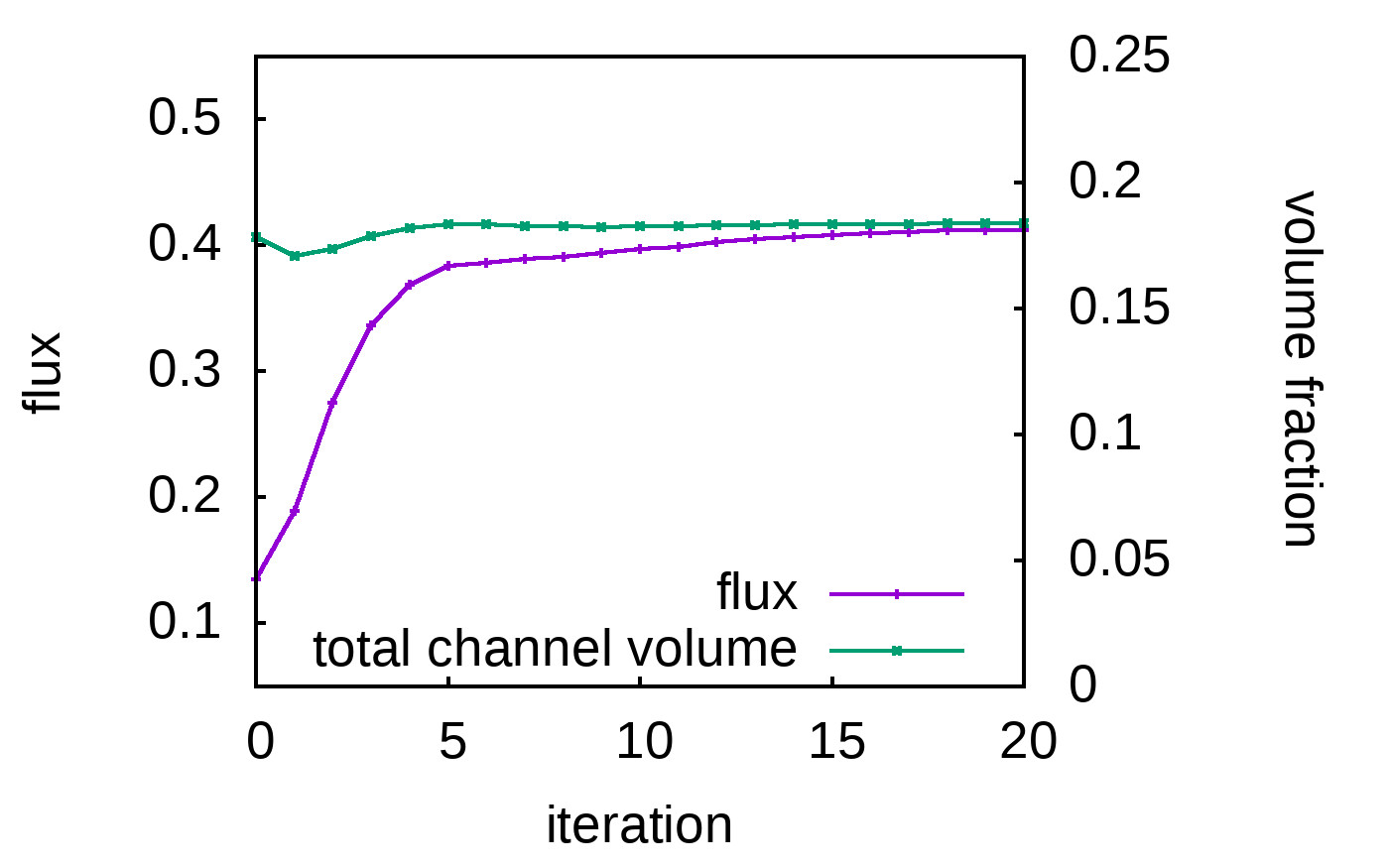}
    \caption{Convergence plots for design shown in \cref{fig:poroel_rot_lambda-10}.}
    \label{fig:poroel_rot_lambda-10-convergence}
\end{figure}

Let us again set $\Lambda_\Phi = 1$ and $\Lambda_\Psi = -10$, as in \cref{fig:min-poroel_no-rot_lambda-10}, and observe in \cref{fig:example-poroel-lambdaPsi-10-iter-1,fig:example-poroel-lambdaPsi-10-optDesign} how the design evolves as both physical models counteract each other: the mechanical model strives for as much material as possible to minimize the compliance while the fluid flux is maximized when there is less material in the design domain. The convergence plot for the merit function $\Jcal$ and compliance function $\Phi$, displayed in \cref{fig:poroel_rot_lambda-10-convergence}, shows that the compliance drops in the first iteration, then increases a bit and finally settles around the value of 27.0. 
In general, we observed in our numerical studies, that the largest design changes occur within a few iterations in the beginning. Afterwards, minor changes are made to further tweak the objective. This behavior shows the good quality of the SGP model and its approximations, described in \cref{sec:sgp}. Let us have a closer look into the intermediate designs shown in \cref{fig:example-poroel-lambdaPsi-10-iter-1}. Again, the initial guess  is neither particularly favorable for the mechanical nor for the fluid flow state. After the first iteration, we see in \cref{fig:example-poroel-lambdaPsi-10-iter-1} that some channels, close to the outflow region, are opened widely and cells closer to the mechanical support were adjusted to have narrower fluid channels to improve the mechanical performance of the design. In comparison to the solution in \cref{fig:min-poroel_no-rot_lambda-10}, where the orientation was fixed, this solution has a 1\% smaller compliance and a fluid flux which is about 47\% higher.

We would like to emphasize that local orientation field looks rather smooth although we have neither applied a stress based warm start for the rotation variable, as proposed by \cite{pedersen1989,norris2006}, nor we have employed a regularization technique. We also can observe that the total number of iterations required did not increase after addition of the additional design degrees of freedom.

We conclude this subsection by presenting a Pareto front for this type of bicriterial weighted sum formulation in \cref{fig:pareto-one-cell}. All optimizations were based on the initial guess that is shown in \cref{fig:init-015-015}. This implies that again, no warm starting technique was employed to proceed from one point to the next on the Pareto curve. Nevertheless a Pareto curve is obtained, in which none of the points is dominated by another one. This again is a hint that the SGP method is able to avoid poor local solutions. The number of outer iterations required to solve the problems corresponding to all points on the Pareto curve varied between $3$ and $31$. The rather low number of $3$ iterations was obtained for the extreme case, where $\Lambda_\Psi=0$.
\pgfplotstableread{pareto_alpha28_phi180.txt}{\tablefine}
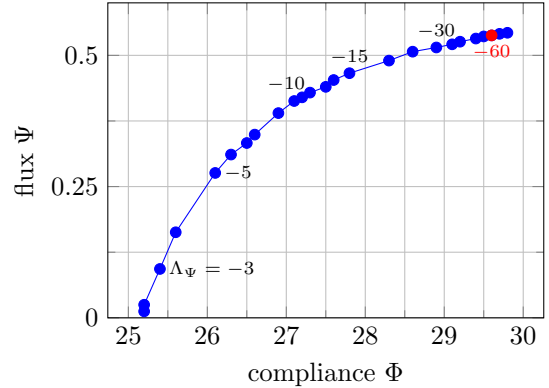
\begin{figure}[htbp]
  \centering
\begin{tikzpicture}
  \begin{axis}[
    ymin=0,
    ymax=0.6,
    xlabel = {compliance $\Phi$},
    ylabel = {flux $\Psi$},
    xtick distance = 1,
    ytick distance = 0.25,
    grid = both,
    minor tick num = 1,
    major grid style = {lightgray},
    width = 0.46\textwidth,
    height = 0.36\textwidth,
    legend cell align = {left},
    legend pos = north west
    ]
        \addplot[blue, mark = *] table [x = {phi}, y = {psi}] {\tablefine};
    \node [right] at (axis cs:  2.54e+01   ,   9.31e-02) {\mbox{\footnotesize$\Lambda_{\Psi}=-3$}};
    \node [right] at (axis cs:  2.61e+01   ,   2.76e-01) {\mbox{\footnotesize$-5$}};
    \node [above] at (axis cs:  2.7e+01   ,   4.13e-01) {\mbox{\footnotesize$-10$}};
    \node [above] at (axis cs:  2.78e+01   ,   4.66e-01) {\mbox{\footnotesize$-15$}};
    \node [above] at (axis cs:  2.89e+01   ,   5.15e-01) {\mbox{\footnotesize$-30$}};
    \node [text=red,below] at (axis cs:  2.96e+01 ,  5.38e-01) {\mbox{\footnotesize$-60$}};
    
    \addplot[
    color=red,
    mark=*,
    ] coordinates {
      (2.96e+01   ,   5.38e-01)
    };
  \end{axis}
\end{tikzpicture}
\caption{Pareto front for varying $\Lambda_\Psi$ in weighted-sum formulation $\Fcal_\text{phys} = \Phi +\Lambda_\Psi \Psi$. The optimization was based on cells of type $1$ and the initial design was always $[0.15,0.15,0]^\nel$. As we are minimizing $\Phi$ and maximizing $\Psi$, a point $P=(P_\Phi,P_\psi)$ in the image space of $\Phi$ and $\Psi$ is dominating a point $Q=(Q_\Phi,Q_\psi)$ if $P_\Phi \leq Q_\Phi$ \emph{and} $P_\Psi \geq Q_\Psi$.}\label{fig:pareto-one-cell}
\end{figure}
The optimized designs for various choices of $\Lambda_\Psi$ are visualized in \cref{fig:rot_one_cell_var_lamb}.
It is observed that the with decreasing  $\Lambda_\Psi$ the compliance minimized is design (\cref{fig:rot_one_cell_var_lamb_a}) is almost smoothly transformed into a fully flux based design (\cref{fig:rot_one_cell_var_lamb_h}).
\begin{figure*}[ht!]
    \centering
    \subfloat[Compliance minimized design\label{fig:rot_one_cell_var_lamb_a}]{\includegraphics[width=0.23\textwidth]{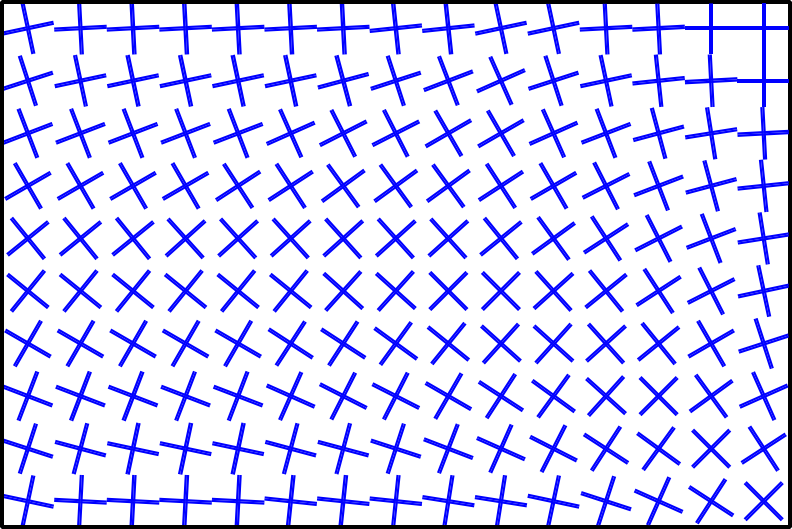}\quad} 
    \subfloat[$\Lambda_\Psi=-3$]{\includegraphics[width=0.23\textwidth]{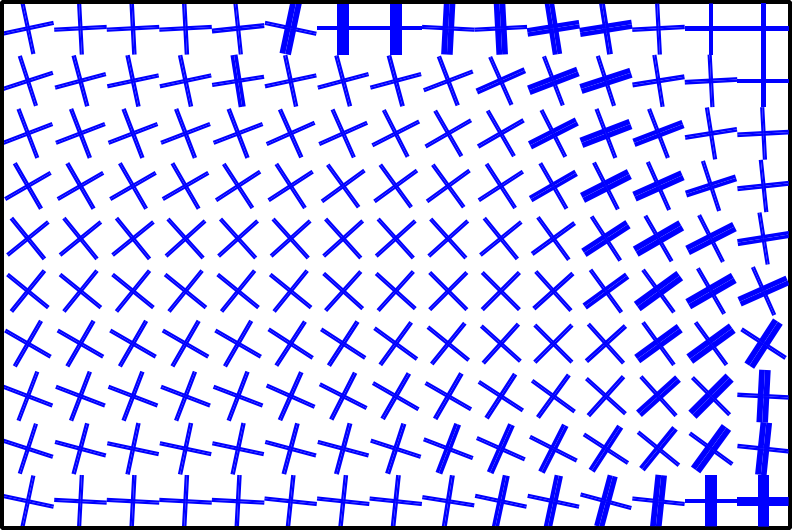}} \quad
    \subfloat[$\Lambda_\Psi=-5$]{\includegraphics[width=0.23\textwidth]{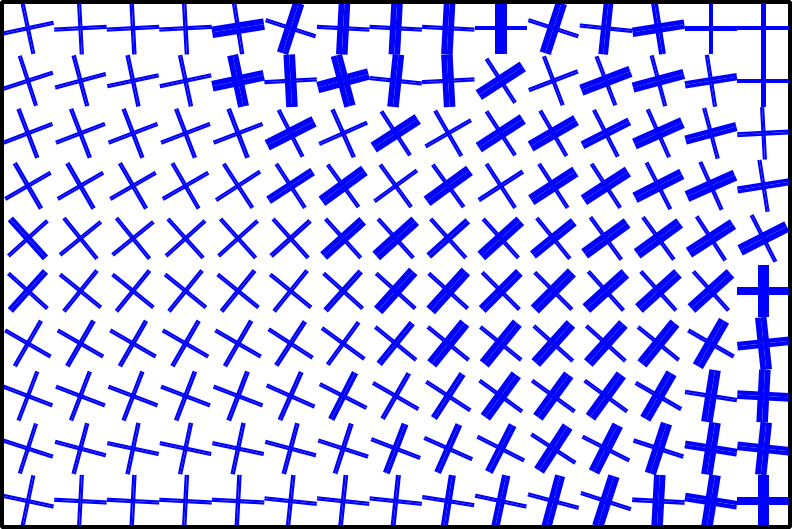}} \quad
    \subfloat[$\Lambda_\Psi=-10$]{\includegraphics[width=0.23\textwidth]{poroel_alpha28_phi180_beta60_15x10x2_init015_phi0_lambda10_optDesign}} \\
    \subfloat[$\Lambda_\Psi=-15$]{\includegraphics[width=0.23\textwidth]{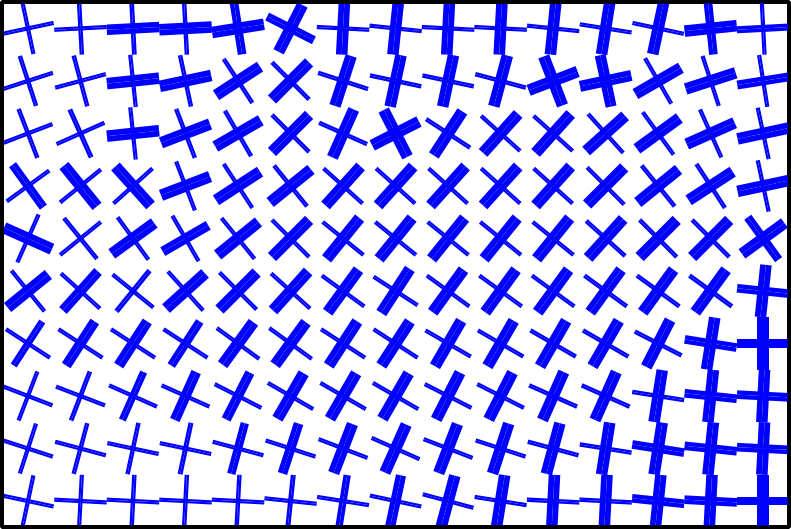}} \quad
    \subfloat[$\Lambda_\Psi=-30$]{\includegraphics[width=0.23\textwidth]{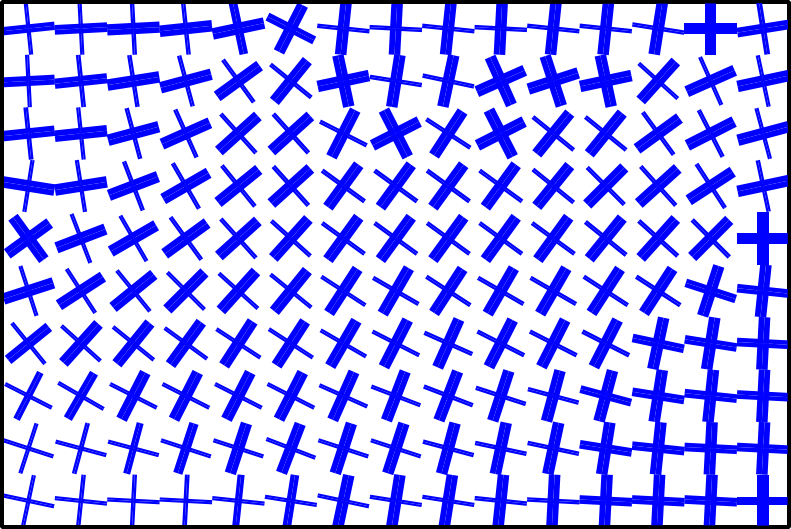}} \quad
    \subfloat[$\Lambda_\Psi=-60$]{\includegraphics[width=0.23\textwidth]{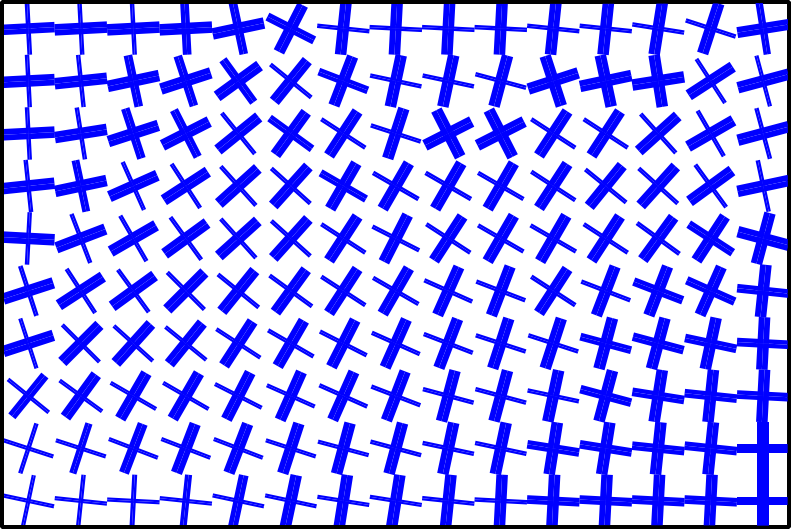}} \quad
    \subfloat[Flux maximized design\label{fig:rot_one_cell_var_lamb_h}]{\includegraphics[width=0.23\textwidth]{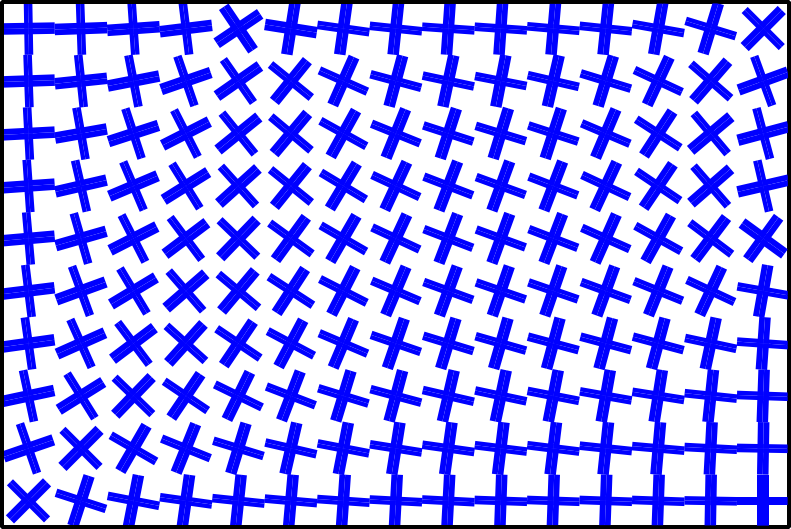}}
    \caption{Visualization of optimized designs associated with the labeled points in \cref{fig:pareto-one-cell}.}
    \label{fig:rot_one_cell_var_lamb}
\end{figure*}

\subsection{Optimization with two unit cell types}\label{sec:two-cells}
\begin{figure}[htbp]
    \centering
    \subfloat[Only cells of type $1$\label{fig:min-compl-M1}]{\includegraphics[width=0.22\textwidth]{poroel_alpha28_phi180_beta60_15x10x2_init015_phi0_lambda0_optDesign_z-0}}\quad
    \subfloat[Cells of type 2 Optimized design with $\Phi_\text{opt}=19.62$\label{fig:min-compl-M2}]{
    \includegraphics[width=0.22\textwidth]{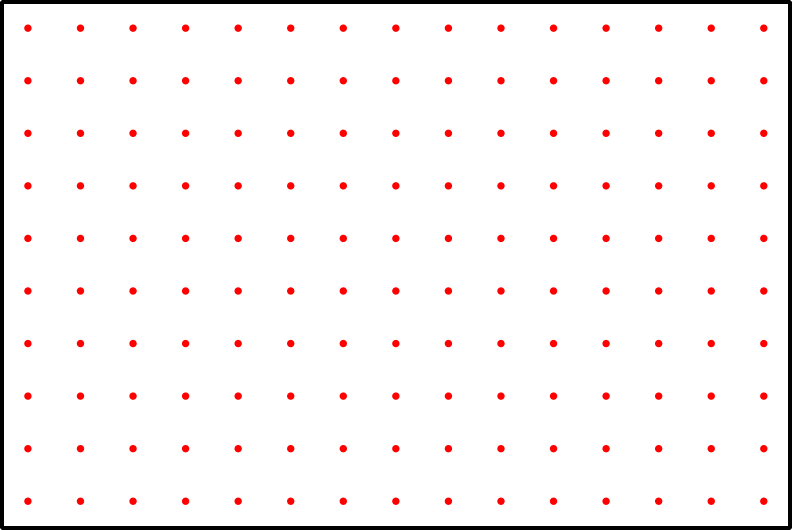}}
    \caption{Compliance minimized designs: \protect\subref{fig:min-compl-M1} only allowing cells of type 1 and \protect\subref{fig:min-compl-M2} allowing choices of type 1 and 2. The red dots visualize the void inclusions of cells of type 2. The optimized compliance of design \protect\subref{fig:min-compl-M2} is 24\% better than compliance of the optimized design \protect\subref{fig:min-compl-M1}.}
    \label{fig:cantilever_no-vol_two-materials}
\end{figure}

We want to study the ability of SGP to handle more than one unit cell type. For this purpose, we add unit cell type $2$ that comprises of a void sphere surrounded by matrix material (see second row of \cref{fig:design_params}). The only design parameter is the radius $r_s \in [0.1,0.4]$ of the void sphere in this case. The smaller the void sphere, the higher the volume fraction of the matrix phase and therefore the stiffer the cell. Thus, cells of type $2$ are particularly favorable for the mechanical part of the objective. When only optimizing the compliance, we obtain the trivial solution shown in \cref{fig:cantilever_no-vol_two-materials}. 

For the fluid flow, cells of type $2$ are futile as they are not permeable. However, for numerical reasons, we set the permeability of the latter cells to 0.001. Cells of type $1$ have orthotropic mechanical properties and transversal isotropic permeability tensors, whereas cells of type $2$ have isotropic mechanical properties and no permeability. Although cell types $1$ and $2$ are disjunct in their parameter spaces, the corresponding ranges of volume fractions, of the stiff matrix material, overlap. We have $\rho\left(\Hcal_1([0.08,0.08])\right) = 87.9\%,\ \rho\left(\Hcal_1([0.22,0.22])\right) = 71.54\%$ and $\rho\left(\Hcal_2(0.4)\right)=73.19\%,\ \rho\left(\Hcal_2(0.1)\right) = 99.6\%$. 
$A^{\text{nodes}}_2$, the basis for the interpolation of $\Hcal_2$, consisted of 30 uniformly distributed samples for $r_s \in [0.1,0.4]$ and the optimization procedure was performed on $A_2^\text{grid}$ with 60 samples, again uniformly distributed. 

Next, we present the updated Pareto front for compliance minimization and fluid flux maximization with both unit cell types in \cref{fig:pareto_two-cells}. 
\pgfplotstableread{pareto_alpha28_phi180_beta60_twoCells.dat}{\paretotwocellsfine}
\pgfplotstableread{pareto_alpha28_phi180.txt}{\tableonecell}
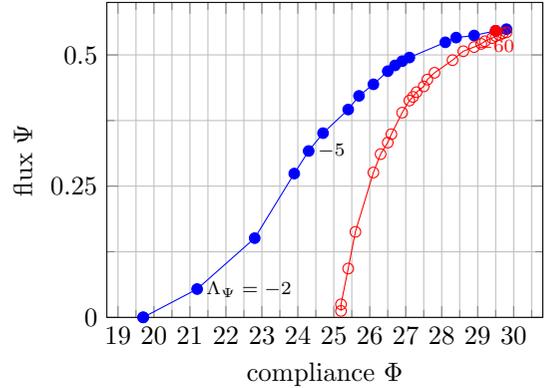
\begin{figure}[h!]
  \centering
\begin{tikzpicture}
  \begin{axis}[
    ymin=0,
    ymax=0.6,
    xlabel = {compliance $\Phi$},
    ylabel = {flux $\Psi$},
    xtick distance = 1,
    ytick distance = 0.25,
    grid = both,
    minor tick num = 1,
    major grid style = {lightgray},
    width = 0.46\textwidth,
    height = 0.36\textwidth,
    legend cell align = {left},
    legend pos = north west
    ]
    \addplot[blue, mark = *] table [x = {phi}, y = {psi}] {\paretotwocellsfine};
    \addplot[red, mark = o] table [x = {phi}, y = {psi}] {\tableonecell};
    \node [right] at (axis cs:  2.12e+01   , 5.40e-02){\mbox{\footnotesize$\Lambda_{\Psi}=-2$}};
    \node [right] at (axis cs:  2.43e+01   ,   3.17e-01) {\mbox{\footnotesize$-5$}};
    \node [text=red,below] at (axis cs:  2.95e+01 ,  5.46e-01) {\mbox{\footnotesize$-60$}};
    
    \addplot[
    color=red,
    mark=*,
    ] coordinates {
      (2.95e+01   ,   5.46e-01)
    };
  \end{axis}
\end{tikzpicture}
\caption{Comparison of Pareto curves for varying $\Lambda_\Psi$. Blue: optimization with cells of type 1 and 2. Red: optimization with only cells of type 1. The blue curve clearly dominates the red curve.}
\label{fig:pareto_two-cells}
\end{figure}
We again stress that we did not use enhanced initial designs for the computation of the points on the Pareto curve. The comparison of the new (blue) curve with the old (red) curve shows that consistently better designs are obtained. Points on the blue curve strictly dominate points on the red curve in the Pareto sense. This is not surprising as, with the addition of a new unit cell type, the design freedom is increased. Still it is worth to mention that the fact that we do not observe any outliers in this respect again underlines the stability of our SGP method. The numbers of required outer iterations varied between 4 and 40, which means that no significant increase in the number of iterations is observed, although a second cell type has been added.
In \cref{fig:opt-two-cells-var-lamb}, we can observe how the number of cells of type 2, in the optimized design, decreases with decreasing $\Lambda_\Psi$. This is expected, as cell type 2 is completely useless for a flux favored design. 
\begin{figure}[ht!]
    \centering
    \subfloat[Compliance minimized design\label{fig:two-cells-min-compl}]{\includegraphics[width=0.23\textwidth]{cantilever_twoCells_init_022_phi_0_optDesign}}\quad
    \subfloat[$\Lambda_\Psi = -2$]{\includegraphics[width=0.23\textwidth]{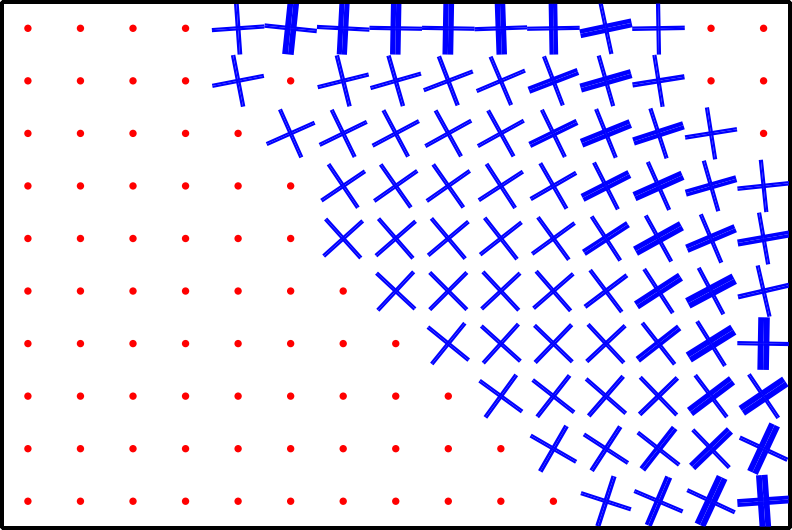}} \quad
    \subfloat[$\Lambda_\Psi = -5$]{\includegraphics[width=0.23\textwidth]{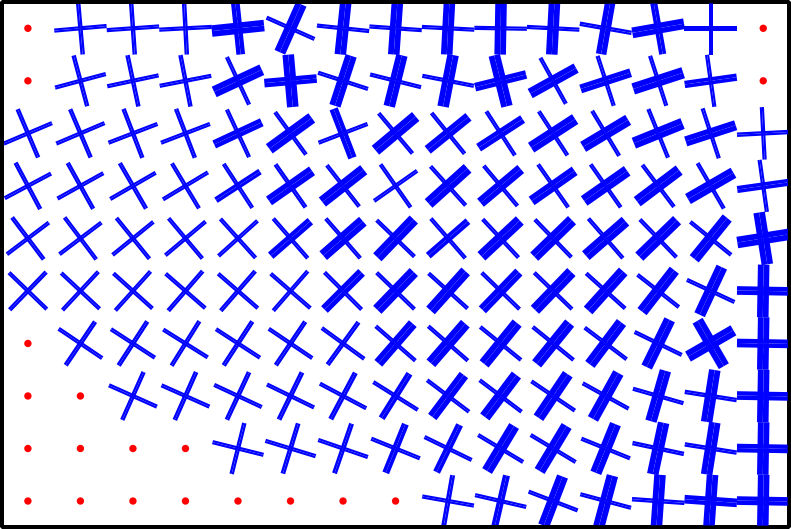}}\\
    \subfloat[$\Lambda_\Psi = -60$]{\includegraphics[width=0.23\textwidth]{poroel_alpha28_phi180_beta60_15x10x2_init015_phi0_lambda60_optDesign}}\quad
    \subfloat[Flux maximized design]{\includegraphics[width=0.23\textwidth]{min_flux_init_008_phi_0_optDesign_z-0}}
    \caption{Results of bicriterial optimization with cells from both type 1 and 2 for varying $\Lambda_\Psi$. The designs visualized here corresponds to the labeled data points of the pareto curve in \cref{fig:pareto_two-cells}.}
    \label{fig:opt-two-cells-var-lamb}
\end{figure}

\begin{figure}[ht]
    \centering
    \subfloat[Optimized design at $z=0$ ]{\includegraphics[width=0.24\textwidth]{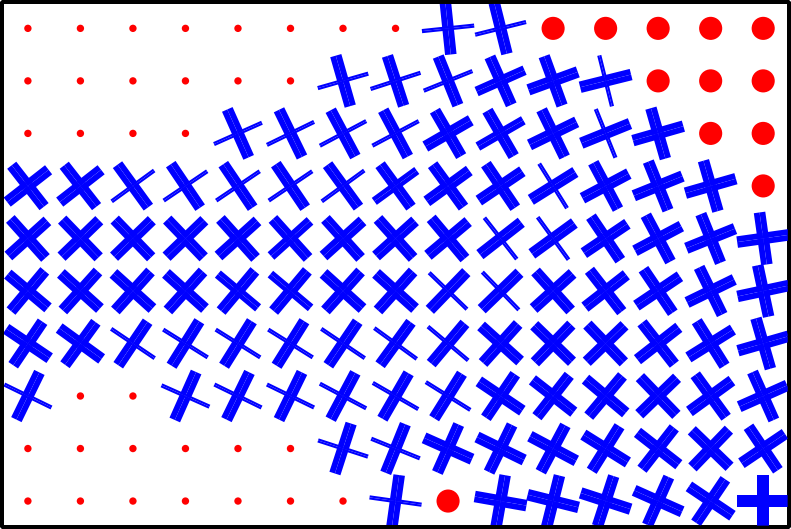}} \\
    \subfloat[Mechanical state with $\Phi_\text{opt}=23.78$]{\includegraphics[width=0.24\textwidth]{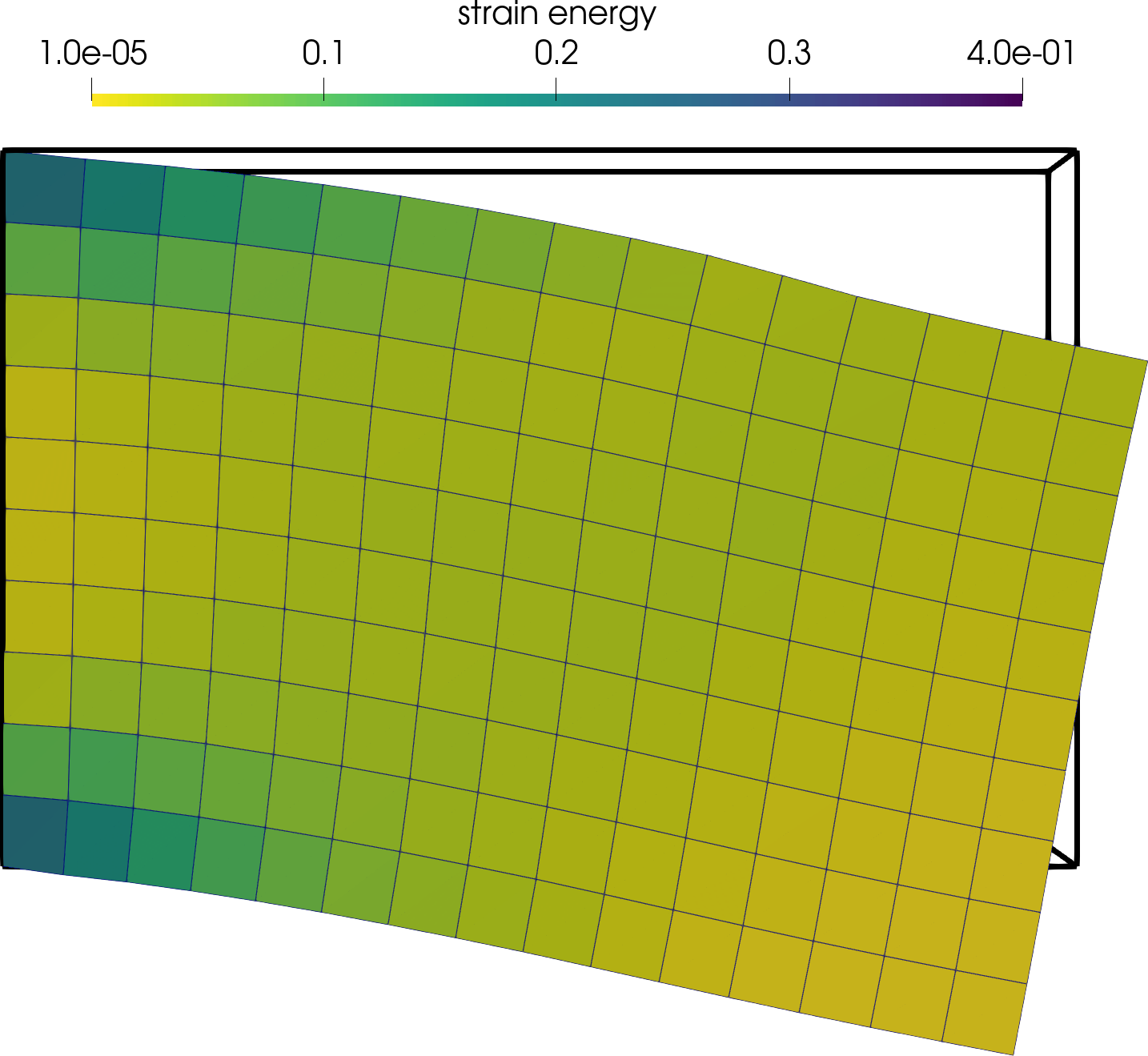}}
    \caption{Result of pure compliance minimization when allowing unit cells of type $1$ and $2$ with an active volume fraction constraint setting $\rhobarm = 0.8$ on the stiff material phase. Comparing to \cref{fig:two-cells-min-compl}, it is observed that only now also cells of type 1 appear in the design. Moreover, the resource constraint leads to a variation of the parameter $r_s$ for cell type 2. }
    \label{fig:cantilever_vol-constr_two-materials}
\end{figure}

We note that so far all results presented have been computed without employing a resource constraint. Just to demonstrate that SGP can also easily handle problems, where a resource constraint is added, we briefly discuss a selected result in \cref{fig:cantilever_vol-constr_two-materials}.
\subsection{Optimization with both cell types and regularization of design labels and interface}\label{sec:two-cells-reg}

We introduce a regularization of the optimization problem by applying a weighted-sum filter $\FF$ (\eg \cite{bruns-filter,bourdin-filter}), that is often used in the context of topology optimization, on regularization labels that are directly related to the unit cells' geometric parameters. For this we introduce mappings
\begin{equation}\label{eq:reg_type_1}
    l_1:
    \begin{cases}
      A_1 &\to \RR^3 \\
      (r_x,r_y,\varphi) &\mapsto R_1
    \end{cases}
\end{equation}
where
\begin{align*}
R_1 &= \left(\frac{r_x-0.08}{0.14}, 
\frac{r_y-0.08}{0.14},
\cos\left(2\frac{\varphi}{\pi}-\frac{\pi}{2}\right)\right)^\top, 
\end{align*}
and
\begin{equation}\label{eq:reg_type_2}
    l_2:
    \begin{cases}
      A_2 &\to \RR^3 \\
      r_s &\mapsto R_2 = (-1,-1,-1)^\top.
    \end{cases}
\end{equation}

This choice of labeling has the following effects: Within type 1, the maximal distance from lower to upper label bound is 1. This is the same distance required to jump from the stiffest cell of type 1, with $r_x=r_y=0.08$, to any cell of type 2. Therefore, the interface between cells of type 1 and 2 is also penalized. The most expensive change is a jump from type 1, which is preferred by the compliance, to any cell of type 2, which is most beneficial for the fluid flux. 
The shifted cosine function appearing in the expression for $(R_1)_3$ is employed to circumvent disambiguities for the angular variable.

\begin{figure}[ht]
    \centering
    \subfloat[Initial design]{\includegraphics[width=0.22\textwidth]{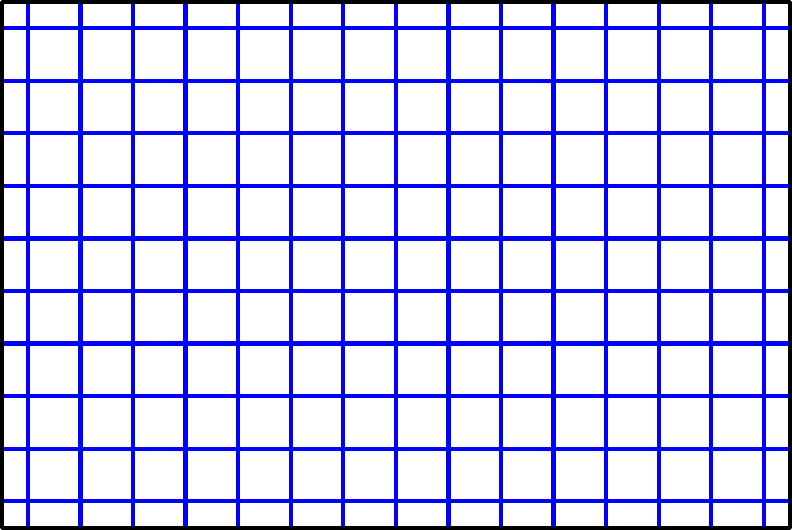}} \quad 
    \subfloat[No regularization\label{fig:opt_two_cells_reg_nofilt}]{\includegraphics[width=0.22\textwidth]{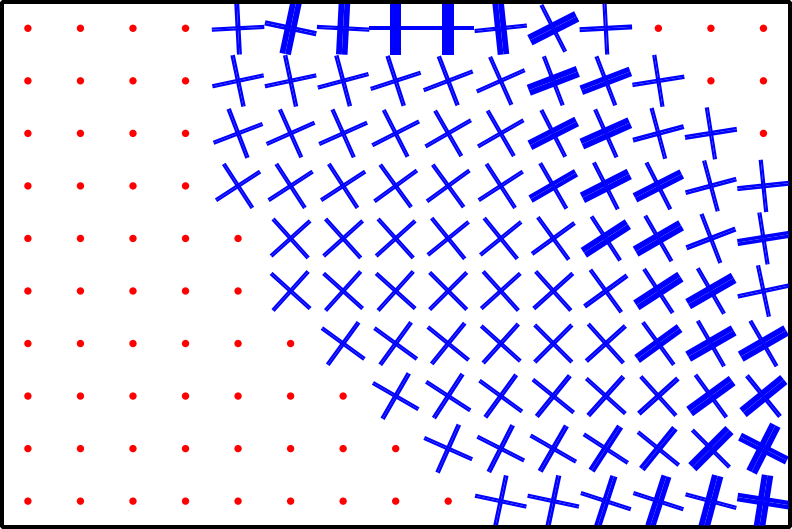}} \quad
    \subfloat[$\LambXi = 0.01$\label{fig:opt_two_cells_reg_p001}]{\includegraphics[width=0.22\textwidth]{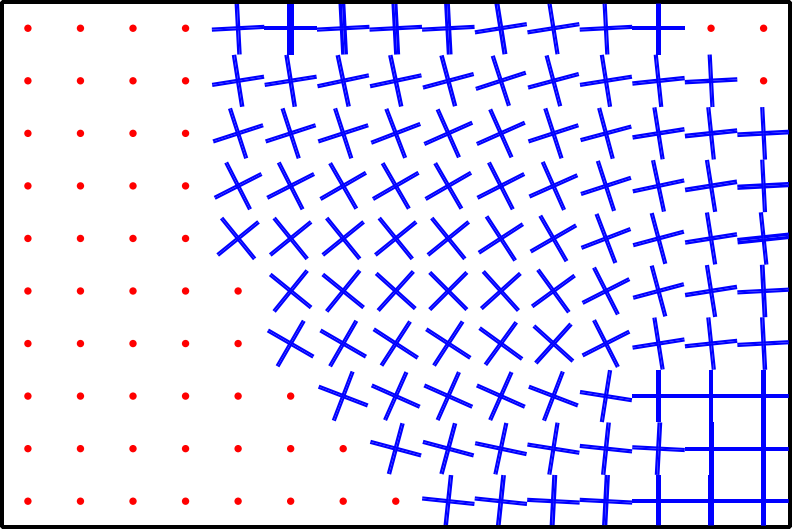}} \\
    \subfloat[$\LambXi = 0.02$]{\includegraphics[width=0.22\textwidth]{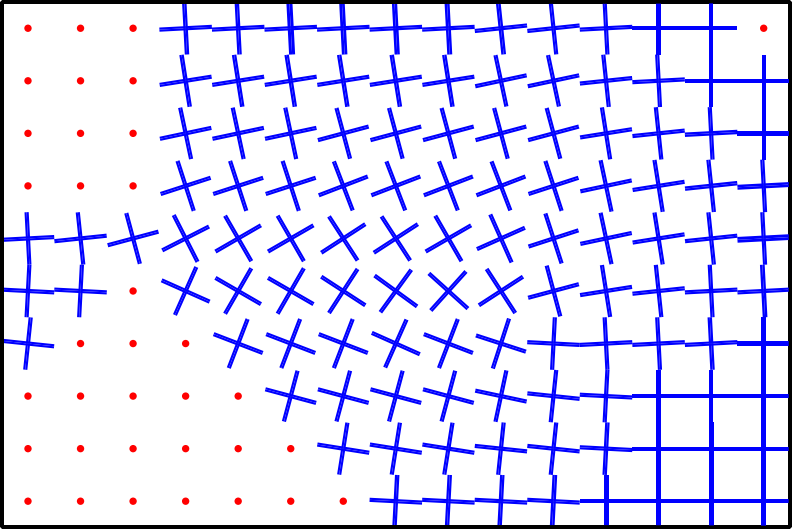}} \quad
    \subfloat[$\LambXi = 0.025$]{\includegraphics[width=0.22\textwidth]{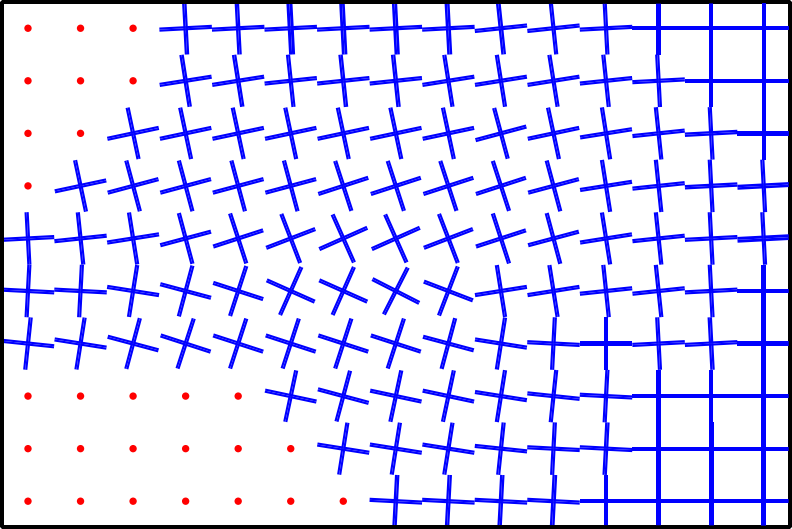}} \quad
    \caption{Results for varying $\LambXi$ with filter radius of $1.3$ elements and $\Lambda_\Psi = -3$.}
    \label{fig:opt_two_cells_reg}
\end{figure}

Employing these regularization labels, $\Jcal_\text{reg}$ from \cref{eq:Jreg-approx} changes to 
\begin{equation}
    \Jreg(\Rb) = \frac{1}{2} \sum_{\ell=1}^3 \|\Rb_\ell - \FF (\Rb_\ell) \|^2,
\end{equation}
where $\Rb_\ell \in \RR^{\nel}$ collects the $\ell$-the components of the regularization label assigned to each finite element, which is  defined by formula \cref{eq:reg_type_1} or \cref{eq:reg_type_2}, if cell type 1 or cell type 2 is chosen for the corresponding finite element $e$, respectively.\\
Next, we study the influence of regularization with the optimized result for the particular choice $\Lambda_\Psi = -3$. The result displayed in \cref{fig:opt_two_cells_reg} displays the changes in design with increasing regularization parameter $p_\text{filt}$. 
The respective objective values are listed in \cref{tab:opt_two_cells_reg}. The regularization of fluid channel radii can be observed well when comparing the designs in the right lower corner of \cref{fig:opt_two_cells_reg_nofilt} and \cref{fig:opt_two_cells_reg_p001}. With increasing $p_\text{filt}$, the interface between unit cell types 1 and 2, at the right upper corner of the design domain, vanishes and the design is dominated by cells of type 1.
\begin{table}[htb]
    \centering
    \begin{tabular}{c||c|c|c|c}
        $\LambXi$ & $\Jcal_\text{mer,opt}$ & $\Jcal_\text{reg,opt}$ & $\Phi_\text{opt}$  & $\Psi_\text{opt}$  \\
        \hline
        0       &21.34    &11.5       &21.57426      &0.0765\\
        0.01    &21.65    &0.0389     &21.65041      &0.0140\\
        0.011   &21.67    &0.0484     &21.66846      &0.0142\\
        0.015   &21.99    &0.0747     &21.95892      &0.0139\\
        0.02    &22.40    &0.092      &22.34912      &0.0135\\
        0.025   &22.70    &0.0712     &22.66730      &0.0136
    \end{tabular}
    \caption{Performance of designs shown in \cref{fig:opt_two_cells_reg} with $\Jcal_\text{mer,opt}(\LambXi) = \Jcal_\text{reg,opt}(\LambXi) + \Phi_\text{opt} + \Lambda_\Psi \Psi_\text{opt}$}
    \label{tab:opt_two_cells_reg}
\end{table}

\section{Conclusion and Outlook}
We presented an Sequential Global Programming (SGP) approach to homogenization-based structural optimization which can be viewed as an free material optimization constrained by the set of admissible geometric material parameters. 

By means of numerical examples, where we successively added more ingredients to the optimization problem, we demonstrated that the proposed SGP approach, with its first-order approximations, provides good and reasonable optimized designs without the necessity of  particular design initialization or the employment of a regularization strategy for purposes of convergence. Furthermore, SGP is able to handle several material classes with disjunct parameter sets without additional interpolation and penalization strategies. We further observed that optimizing the local orientation of the microstructure brings along a significant improvement, up to 48\%, of the fluid flux. We have not actively addressed the subject of connectivity within the microstructure, that is to ensure connectivity of the fluid saturated channels. However, the regularization approach presented in \cref{sec:two-cells-reg} can be used to control the degree of variation of the local microstructure rotation and we have seen, by means of the presented numerical examples, that only a mild regularization has already a fair impact on the design.

Although the resolution of the finite element approximation, and thus the number of design elements, of the examples in section \cref{sec:numerical-results} was chosen rather coarsely, it served the purpose of demonstrating the presented features of SGP. With regard to finer resolutions: the algorithm can be well parallelized with respect to the design elements due to the block-separability of the first-order approximations. 

The brute-force approach in the subproblem solver, described in \cref{alg:sub-problem}, can further be speeded up by employing a hierarchical scanning of the design grids $\Agrid_i$: Start with a rather coarse number of samples and determine the minimizer among those. In the next level, consider only the current minimizer and its neighbors and perform the same search within this subset of $\Agrid_i$, for all $i \in I$. Repeat this step until the maximum desired number of levels or some accuracy is achieved. Note that, with this strategy, the quality of the design depends on the number of samples on the coarsest grid level. An alternative would be to apply a Lipschitz optimization solver, see \cite{Hansen1995}, to each design element and type in a black box manner.

Further research will focus on extending the SGP approach for homogenization-based optimization to transient problems and, in particular, to dynamic metamaterial design. Another challenge is to extend the proposed optimization approach for an approximate treatment of nonlinear two-scale problems with the homogenized coefficients depending on the macroscopic response by virtue of the sensitivity analysis as discussed in \cite{Rohan-Lukes-2015}.

\section{Acknowledgments}
The authors B. N. Vu and M. Stingl gratefully acknowledge the financial support by the German Federal Ministry for Economic Affairs and Climate Action (BMWK) in the course of the FIONA (LuFo VI-1, FKZ: 20W1913F) project. The research conducted by E. Rohan and V. Luke\v{s} was supported by the grant projects GACR 19-04956S and GACR 22-00863K of the
Czech Scientific Foundation.

\section{Statements and Declarations}
The authors declare that they have no known competing financial interests or personal relationships that could have appeared to influence the work reported in this paper.

\section{Replication of results}
The algorithm of the proposed optimization approach was described in \cref{alg:sgp-multimat} and \cref{alg:sub-problem}. Its implementation, as well as exemplary problem settings and respective data to reproduce  the numerical results presented in \cref{sec:numerical-results}, are publicly available on \href{https://gitlab.com/bnvu/sgp-poroel}{https://gitlab.com/bnvu/sgp-poroel}.

\bibliography{main}
\end{document}